\documentclass[useAMS,usenatbib,usegraphicx]{mn2e} 
\usepackage{amsfonts,amsmath,amssymb}
\usepackage{times}

\newcommand{\nb}{$N$-body} 
\newcommand{\lcdm}{$\Lambda$CDM} 
\newcommand{\n}{\noindent}

\newcommand{\tsl}{\hbox{$T_\mathrm{sl}$}}
\newcommand{\lx}{\hbox{$L_\mathrm{X}$}}
\newcommand{\lxt}{\hbox{$L_\mathrm{X}$-$T$}}
\newcommand{\lxtsl}{\hbox{$L_\mathrm{X}$-$T_{\rm sl}$}}

\newcommand{\mt}{\hbox{$M$-$T$}}

\newcommand{\tm}{\hbox{$T$-$M$}}
\newcommand{\tslm}{\hbox{$T_{\rm sl}$-$M$}}

\newcommand{\lxm}{\hbox{$L_\mathrm{X}$-$M$}}

\newcommand{\yx}{\hbox{$Y_\mathrm{X}$}}
\newcommand{\yxm}{\hbox{$Y_\mathrm{X}$-$M$}}
\def\Msun{\hbox{$\rm\, M_{\odot}$}}
\def\Zsun{\hbox{$\rm\, Z_{\odot}$}}

\begin{document}

\title[Evolution of cluster scaling relations]{The evolution of galaxy cluster X-ray scaling relations} \author[C.~J.~Short et al.]{C.~J.~Short,$^1$\thanks{E-mail: C.Short@sussex.ac.uk} P.~A.~Thomas,$^1$ O.~E.~Young,$^1$ F.~R.~Pearce,$^2$ A.~Jenkins$^3$ and \newauthor O.~Muanwong$^4$\\ $^1$Astronomy Centre, University of Sussex, Falmer, Brighton, BN1 9QH, United Kingdom \\ $^2$ Department of Physics and Astronomy, University of Nottingham, Nottingham, NG7 2RD, United Kingdom \\ $^3$Institute for Computational Cosmology, Department of Physics, University of Durham, South Road, Durham, DH1 3LE, United Kingdom \\ $^4$Department of Physics, Khon Kaen University, Khon Kaen, 40002, Thailand}
\maketitle

\begin{abstract}
We use numerical simulations to investigate, for the first time, the joint effect of feedback from supernovae (SNe) and active galactic nuclei (AGN) on the evolution of galaxy cluster X-ray scaling relations. Our simulations are drawn from the Millennium Gas Project and are some of the largest hydrodynamical \nb\ simulations ever carried out. Feedback is implemented using a hybrid scheme, where the energy input into intracluster gas by SNe and AGN is taken from a semi-analytic model of galaxy formation. This ensures that the source of feedback is a population of galaxies that closely resembles that found in the real Universe. We show that our feedback model is capable of reproducing observed local X-ray scaling laws, at least for non-cool core clusters, but that almost identical results can be obtained with a simplistic preheating model. However, we demonstrate that the two models predict opposing evolutionary behaviour. We have examined whether the evolution predicted by our feedback model is compatible with observations of high-redshift clusters. Broadly speaking, we find that the data seems to favour the feedback model for $z\lesssim 0.5$, and the preheating model at higher redshift. However, a statistically meaningful comparison with observations is impossible, because the large samples of high-redshift clusters currently available are prone to strong selection biases. As the observational picture becomes clearer in the near future, it should be possible to place tight constraints on the evolution of the scaling laws, providing us with an invaluable probe of the physical processes operating in galaxy clusters.
\end{abstract}


\section{Introduction}

Galaxy cluster surveys are a potentially powerful means of placing tight constraints on key cosmological parameters, independent of other methods such as the measurement of cosmic microwave background anisotropies. This is primarily because the mass function of galaxy clusters is highly sensitive to different choices of model parameters. It is therefore essential to determine how the mass function varies with redshift if we are to exploit clusters as a cosmological probe. This is not trivial since the total masses of galaxy clusters must first be inferred from their observable properties. 

Mass estimates can be derived from X-ray observations of the hot, diffuse intracluster medium (ICM) by assuming that the gas is in hydrostatic equilibrium within the cluster gravitational potential well (e.g. \citealt{SAR88}). However, this requires the accurate determination of gas density and temperature profiles out to large radii. Although this has become common practice for low-redshift clusters with the advent of \emph{Chandra} and \emph{XMM-Newton}, measuring high-quality profiles of distant clusters requires long observation times, so hydrostatic mass estimates have only been made for a small number of high-redshift clusters. Furthermore, the assumption of hydrostatic equilibrium only applies to dynamically relaxed systems, so this technique cannot be applied to unbiased cluster samples.

In cases where a hydrostatic estimate is not possible, cluster masses can be inferred from the relationships that exist between X-ray observables, like luminosity, \lx, temperature, $T$, and the total mass, $M$. These scaling relations are predicted by the simple self-similar model of cluster formation, where the ICM is heated solely by gravitational processes, such as adiabatic compression and shocks induced by supersonic accretion \citep{KAI86}. However, the observed \lxm\ relation is steeper than expected from gravitational heating alone (e.g. \citealt{REB02,CRB07,VBE09,PCA09}), as is the \lxt\ relation (e.g. \citealt{MAR98,ARE99,WXF99,EDM02,PCA09}). This departure from self-similarity is due to an excess of entropy\footnote{We define entropy as $K=k_{\rm B}T/n_\mathrm{e}^{\gamma-1}$, where $k_{\rm B}$ is Boltzmann's constant, $n_\mathrm{e}$ is the electron number density and $\gamma=5/3$ is the ratio of specific heats for a monoatomic ideal gas.} in cluster cores \citep{PCN99,LPC00,FJB02}. This extra entropy prevents gas from being compressed to high densities, reducing its X-ray emissivity compared to the self-similar prediction. The effect will be more pronounced in galaxy groups since they have shallower potential wells, leading to a steepening of the \lxm\ relation (and \lxt\ relation) as desired. The main physical processes thought to be responsible for boosting the entropy of the ICM are heating from astrophysical sources, such as supernovae (SNe) and active galactic nuclei (AGN), and the removal of low-entropy gas via radiative cooling (see \citealt{VOI05} for a review).

Unfortunately, scaling relations are not a perfect means of converting from X-ray observables to mass because of their intrinsic scatter. The dominant source of scatter about relations involving \lx\ is the dense, highly X-ray luminous central regions of cool core (CC) clusters (e.g. \citealt{OMB06,CRB07}). One way of reducing this scatter is simply to exclude the core region from the measurements (e.g. \citealt{MAR98}). Another source of scatter is cluster mergers, which can cause clusters to shift (approximately) along the \lxt\ relation, but away from the $M$-$T$ relation \citep{RTK04,KVN06}. However, \citet{KVN06} recently demonstrated that the quantity \yx, defined as the product of the gas mass $M_{\rm gas}$ and the X-ray temperature, is extremely tightly correlated with total mass and is insensitive to cluster mergers. This result has been confirmed by independent numerical simulations \citep{PBM07} and several observational studies \citep{APP07,MAU07,VBE09}, suggesting that reliable cluster mass estimates can indeed be derived from simple observables.

Given that scaling relations enable us to estimate the masses of clusters with poor quality X-ray data, it is clearly vital to ensure that they are well calibrated over a wide redshift range, so we can improve cosmological constraints derived from the mass function. Furthermore, measuring the evolution of X-ray scaling laws should reveal information on the nature of the non-gravitational heating and cooling processes responsible for shaping galaxy clusters \citep{MKT06}.

At low redshift, scaling relations are reasonably well calibrated, at least for relaxed clusters, since reliable hydrostatic mass estimates are readily available. Measuring these relations at high-redshift is considerably more challenging, with fewer results available in the literature. As will become clear from the following brief review, the observational picture is far from settled at present. 

\citet{ETB04} used a sample of $28$ galaxy clusters with $0.4<z<1.3$ to measure the $M$-$T$, \lxm\ and \lxt\ relations. Upon comparing their high-redshift $M$-$T$ relation with local counterparts, they found that the normalisation evolved in a manner well described by the self-similar model. By contrast, the normalisation of both the high-redshift \lxm\ and \lxt\ relations was lower than expected from self-similar scaling arguments. Negative evolution\footnote{Throughout this work, we consider the evolution of scaling relations relative to that predicted by the self-similar model. Given a particular scaling law, we say there is negative (positive) evolution if the normalisation at high-redshift is lower (higher) than anticipated from self-similar scaling.} of the \lxt\ relation was also reported by \citet{HCS07}, based on \emph{XMM-Newton} observations of a single cluster at $z=1.457$. Another study was performed by \citet{MJE06} who used a sample of $11$ clusters in the redshift range $0.6<z<1$. They concluded that there is no evidence for any evolution of the $M$-$T$, \lxm\ and \lxt\ relations beyond that anticipated from self-similar theory. Similar results for the \lxt\ relation were also found by \citet{VVM02} and \citet{LBR04}. However, such analyses often rely on simplifying assumptions, like isothermal temperature profiles and $\beta$-model profiles for the gas density, which can lead to large biases in the mass determination (e.g. \citealt{MAV97}). 

A different approach was adopted by \citet{MEM07}, whose sample consisted of $24$ clusters in the redshift range $0.14<z<0.82$. For each object in their sample, they used the measured gas density profile and the equation of hydrostatic equilibrium to determine the best-fitting temperature profile by varying the free parameters in the assumed model for the dark matter density profile. In conflict with most of the results discussed above, they reported slight negative evolution of the \mt\ and \lxm\ relations, and positive evolution of the \lxt\ relation.

More recently, small samples of distant clusters ($\sim 10$ objects) with hydrostatic mass estimates have been used to measure the \yxm\ \citep{MAU07}, $M$-$T$ \citep{KOV05,KOV06} and \lxt\ \citep{KOV05} relations at high-redshift ($z\sim 0.7$). The evolution of the normalisation of the \yxm\ and $M$-$T$ relations was again found to be consistent with the self-similar model. On the other hand, the \lxt\ relation exhibited positive evolution, consistent with the results of \citet{MEM07}, but not the negative evolution measured by \citet{ETB04} and \citet{HCS07}. Beyond $z\sim 0.7$, there are very few clusters with a hydrostatic mass estimate; a notable exception is the $z=1.05$ object discovered by \citet{MJP08}, who demonstrated that the position of this cluster on the high-redshift \yxm, $M$-$T$ and \lxm\ relations can be explained by self-similar scaling arguments. 

Although several observational studies indicate that the self-similar model provides an adequate description of the evolution of X-ray scaling relations, \citet{VOI05} suggests that self-similar evolution cannot persist to arbitrarily high redshift, because of the increasing importance of radiative cooling and feedback from galaxy formation. An example of an observational result that supports this argument is that of \citet{BGF07}. Using a sample of $39$ distant clusters ($0.25<z<1.3$), they found that the evolution of the \lxt\ relation is comparable with the self-similar prediction for $0<z<0.3$, but negative at higher redshifts.

One reason for the lack of concordance between different observational studies is inconsistent cluster selection. With current heterogeneous archival cluster samples, it is very difficult to account for selection biases, which may vary from sample to sample and can mimic evolution. An attempt to quantify the impact of selection effects on the evolution of scaling relations was made by \citet{PPA07}, who focused on the \lxt\ relation. From an analysis of their raw data, they found evidence for non-monotonic evolution, at odds with the self-similar prediction. However, after taking selection effects into account, they found that the evolution of the \lxt\ relation was consistent with self-similar scaling, although this was a tentative result because of poor statistics. Nevertheless, the salient point to take from their work is that modelling the full source-selection process can drastically alter the measured evolution, so we must seriously question any attempt to assess the evolution of X-ray scaling laws that does not attempt to do this. In more recent work, \citet{MAR09,MAE09} have developed a method for rigorously accounting for selection effects in order to constrain the scaling relations at low and high redshifts, including their evolution. They concluded that the \tm\ and \lxm\ relations both show no departures from self-similar evolution up to $z\approx 0.5$.

In the near future, the \emph{XMM Cluster Survey} (XCS; \citealt{RVL01}) will provide a complete sample of clusters spanning a broad redshift range, $0<z\lesssim 1.5$, all coming from the same survey and selected with well-defined criteria. Since the survey selection function will be known, the evolution of X-ray scaling relations will be measured with unprecedented accuracy out to high-redshift. Cluster surveys based on the Sunyaev-Zel'dovich (SZ) effect, such as those being conducted with the \emph{South Pole Telescope} \citep{VCD10} and the \emph{Atacama Cosmology Telescope} \citep{HAA09}, are also expected to produce large catalogues of clusters covering a wide redshift range with a well-defined mass threshold. Like XCS, these samples will be ideal for studying the evolution of cluster scaling relations.

There are also several other reasons why results from different studies may appear to be contradictory. For example, the measured evolution can be affected by: poor statistics due to small sample sizes; the local scaling relation chosen to compare high-redshift data with; different definitions of evolution (i.e. whether the expected self-similar behaviour is scaled out first or not); different choices of scale radius used to define a cluster (i.e. redshift-dependent or not); and whether the core region is excised or not (and the size of the core region). In short, extreme care is required when comparing observational results with each other, and with theoretical results from numerical simulations.

Theoretical studies of cluster formation have mainly concentrated on explaining the lack of self-similarity inherent in low-redshift scaling relations, rather than the evolution of scaling laws. To this end, numerous mechanisms for raising the entropy of the intracluster gas have been implemented in hydrodynamical \nb\ simulations, such as radiative cooling (e.g. \citealt{BRY00,PTC00,MTK01,MTK02,DKW02,VBB02,WUX02}), preheating (e.g. \citealt{BEM01,BRM01,MTK02,BGW02,TBS03,BFK05}), star formation and associated feedback from SNe (e.g. \citealt{VAL03,KAY04,KTJ04,KDA07,NKV07,BMS04,BFK05,RSP06}), and black hole growth with associated feedback from AGN (e.g. \citealt{PSS08,MSP10,FBT10}). 

Simulations employing such models have indeed proved capable of successfully reproducing the observed scaling relations for local clusters, at least on average, although often at the expense of excessive star formation. However, we would not expect the thermal history of the ICM to be the same in each case, so we should see differences in evolutionary behaviour. This was first demonstrated by \citet{MKT06} who traced the evolution of the cluster population to $z=1.5$ using three different models: a cooling-only model, a model with preheating and cooling, and a stellar feedback model that self-consistently heats cold gas in proportion to the local star formation rate. While all three schemes produce indistinguishable \lxt\ relations at $z=0$, they predict strongly positive, mildly positive and mildly negative evolution of the \lxt\ relation, respectively. Therefore, given observational data of sufficient quality, it should be possible to rule out unsuitable theoretical models of non-gravitational heating. This emphasises the importance of accurately measuring cluster scaling relations over a range of redshifts.

Other attempts to study the evolution of X-ray scaling laws with numerical simulations are scarce. \citet{EBM04} used a simulation including radiative cooling, star formation and supernova feedback to follow the evolution of the cluster population between $z=1$ and $z=0.5$. They measured a small, but significant, negative evolution in the normalisation of the \lxm\ and \lxt\ relations, and a marginally negative evolution of the $M$-$T$ relation. These results were found to be consistent with the observational data of \citet{ETB04}. \citet{KDA07} used a similar simulation, but with a different prescription for cooling and stellar feedback, to investigate the evolution of the $M$-$T$ and \lxt\ relations in the redshift interval $0<z<1$. The results they obtained are qualitatively the same as those of \citet{EBM04}. However, we note that a rigorous comparison between the results of \citet{EBM04}, \citet{MKT06} and \citet{KDA07} is not possible owing to differences in their data analysis procedures (different definitions of cluster scale radii, etc.)

In this paper, we reconsider the evolution of cluster scaling relations from a theoretical perspective. Our basic goal is to investigate how scaling laws evolve when feedback from AGN is included in numerical simulations, in addition to stellar feedback. As far as we are aware, this is the first time this has been attempted. 

The simulation we use is drawn from the Millennium Gas Project, a series of hydrodynamical \nb\ simulations with the same volume ($500^3 h^{-3}$ Mpc$^3$) and the same initial perturbation amplitudes and phases as the Millennium Simulation \citep{SWJ05}. Feedback from galaxies is incorporated in our simulation via the hybrid approach of \citet{SHT09}, where the energy imparted to the ICM by SNe and AGN is computed from a semi-analytic model (SAM) of galaxy formation. Since SAMs are tuned to reproduce the properties of observed galaxies, the source of feedback in our simulation is guaranteed to be a realistic galaxy population. By contrast, fully self-consistent simulations with radiative cooling, star formation and supernova feedback typically predict that too much gas cools and forms stars (e.g. \citealt{BMS04,KDA07}). We assess how feedback from galaxy formation affects the evolution of scaling relations by comparing our results with those obtained from two other Millennium Gas simulations. The first of these is a control model, where the gas is heated by gravitational processes only. The second includes additional radiative cooling and uniform preheating at high redshift as a simple model of non-gravitational heating from astrophysical sources.

The large volume of the Millennium Gas simulations enables us to resolve statistically significant numbers of clusters at all redshifts relevant to cluster formation. In fact, our cluster samples are some of the largest ever extracted from numerical simulations. For example, in the preheating plus cooling run we have over $20$ times more objects with $T>2\;{\rm keV}$ at $z=0$ and $z=1$ than \citet{KDA07}. Furthermore, we can capture the formation of the richest systems, allowing us to probe the same dynamic range as the observations. The Millennium Gas simulations thus provide an ideal tool for studying the evolution of the cluster population.

The layout of this paper is as follows. In Section \ref{sec:scalerel} we briefly review the self-similar model of cluster formation. The three Millennium Gas simulations are discussed in Section \ref{sec:method}, along with a description of our method for generating cluster catalogues and profiles. In Section \ref{sec:results} we compare results for our three models, starting with a discussion of cluster profiles and X-ray scaling relations at $z=0$, then investigating how profiles and scaling laws evolve from $z=1.5$ to $z=0$. Wherever possible we attempt to place observational constraints on our models. We summarise our results and conclude in Section \ref{sec:conc}.  

\section{Cluster scaling theory}
\label{sec:scalerel}

The simplest model of galaxy cluster formation is that they form via the gravitational collapse of the most overdense regions in the dark matter distribution, and the cluster baryons are heated only by gravitational processes (compression and shock heating) during the collapse. Since non-linear gravitational processes do not introduce any characteristic scale, we would then expect clusters to be self-similar, i.e. scaled versions of each other. With the additional assumptions that clusters are spherically symmetric systems and that the intracluster gas is in hydrostatic equilibrium with the underlying dark matter potential, it is straightforward to derive simple self-similar scaling relations between cluster properties \citep{KAI86}. 

Defining $r_{\Delta}$ as the radius of a spherical volume within which the mean matter density is $\Delta$ times the critical density at redshift $z$, we find that the total enclosed mass, $M_{\Delta}$,\footnote{$M_{\Delta}=4\pi r_{\Delta}^3\Delta\rho_{\rm cr}(z)/3$, where $\rho_{\rm cr}(z)=3H_{0}^2E(z)^2/8\pi G$ is the critical density and $E(z)^2=\Omega_{\rm m,0}(1+z)^3+\Omega_{\Lambda,0}$ in a spatially-flat \lcdm\ cosmological model.} scales with gas temperature, $T_{\Delta}$, as
\begin{equation}
\label{eq:TM}
E(z)^{-2/3}T_{\Delta}\propto M_{\Delta}^{2/3}.
\end{equation}
Under the further assumption that the X-ray emission of the ICM is primarily thermal Bremsstrahlung radiation (which is valid for $T>2$ keV), the luminosity, $L_{{\rm X},\Delta}$, within $r_{\Delta}$ is given by
\begin{equation}
\label{eq:LT}
E(z)^{-1}L_{{\rm X},\Delta}\propto T_{\Delta}^{2}.
\end{equation}
The scaling between X-ray luminosity and total mass follows upon combining equations (\ref{eq:TM}) and (\ref{eq:LT}): 
\begin{equation}
\label{eq:LM}
E(z)^{-7/3}L_{{\rm X},\Delta}\propto M_{\Delta}^{4/3}.
\end{equation}

The quantity $\yx=M_{\rm gas}T$ has recently attracted much attention since it has been shown to be a low-scatter mass proxy, regardless of cluster dynamical state (e.g. \citealt{KVN06}). This is primarily because \yx\ approximates the total thermal energy of the ICM, which is not strongly affected by cluster mergers \citep{PBM07}, unlike \lx\ or $T$ \citep{RIS01}. The self-similar scaling of \yx\ within $r_{\Delta}$ with total mass is  
\begin{equation}
\label{eq:YM}
E(z)^{-2/3}Y_{{\rm X},\Delta}\propto M_{\Delta}^{5/3}.
\end{equation}

The density contrast $\Delta$ governs the scale radius within which one measures the mass of a cluster. The most common choice is to set $\Delta=500$, since $r_{500}$ is the effective limiting radius for reliable observations from \emph{Chandra} and \emph{XMM-Newton}. Throughout this paper we will adopt $\Delta=500$, independent of redshift. 

Note that there are other ways in which the scale radius of a cluster can be defined. In the original model of \citet{KAI86}, clusters were assumed to be self-similar with respect to the mean matter density of the Universe, rather than the critical density. In this case, the $E(z)$ factors in equations (\ref{eq:TM})--(\ref{eq:YM}) are replaced by $(1+z)$. Another possibility is to assume a redshift-dependent density contrast $\Delta(z)$ that is proportional to $\Delta_{\rm vir,z}$, where $\Delta_{\rm vir}(z)\rho_{\rm cr}(z)$ is the mean density within the cluster virial radius (e.g. \citealt{ETB04,MJE06,BGF07}). The value of $\Delta_{\rm vir}(z)$ is taken from the analytical solution for the collapse of a spherical top-hat perturbation, under the assumption that the cluster has just virialised at that redshift (e.g. \citealt{PEE80}). With this choice of density contrast, the $E(z)$ factors in equations (\ref{eq:TM})--(\ref{eq:YM}) become $E(z)\Delta(z)^{1/2}$ instead. Therefore, predictions for the evolution of the cluster population clearly depend on how the scale radius is defined, so caution must be exercised when comparing the results of different observational and theoretical studies. 

Observations of clusters in the local Universe have established that scaling relations between cluster properties do indeed exist, but their form is found to be different to the self-similar predictions. This is because non-gravitational effects, such as radiative cooling and feedback from galaxy formation, modify the entropy structure of the ICM. Lower-mass systems are affected more than massive objects, breaking the self-similarity of the scaling laws. Departures from self-similarity thus provide us with a probe of the non-gravitational processes operating in clusters. Although the self-similar model cannot explain the observed form of scaling relations, there is some observational evidence that it is capable of describing their evolution.

\section{Simulations: The Millennium Gas Project}
\label{sec:method}

In this work we use simulations taken from the Millennium Gas Project, a programme to add gas to the dark matter-only Millennium Simulation \citep{SWJ05}. We present a new addition to the Millennium Gas suite in which feedback is directly tied to galaxy formation, enabling us to investigate how energy input from both SNe and AGN affects the evolution of cluster scaling relations. Hereafter, we refer to this simulation as the FO run. The feedback model adopted is the hybrid scheme of \citet{SHT09}, where a SAM is used to calculate the energy transferred to the intracluster gas by SNe and AGN. An immediate benefit of this approach is that feedback is guaranteed to originate from a galaxy population whose observational properties agree well with those of real galaxies. This is generally not the case in fully self-consistent hydrodynamical simulations that include radiative cooling and stellar feedback because too much gas cools out of the hot phase, leading to excessive star formation (e.g. \citealt{BMS04,KDA07}). It is widely thought that additional heating from AGN is the natural solution to this overcooling problem. Indeed, \citet{MSP10} and \citet{FBT10} have demonstrated that including AGN feedback in hydrodynamical simulations can successfully balance radiative cooling in galaxy groups. However, the stellar fraction is still found to be $\sim 2-3$ times larger than observed in massive clusters \citep{FBT10}. 

By coupling a SAM to a hydrodynamical simulation, \citet{SHT09} were able to reproduce the observed mean \lxt\ relation for groups and poor clusters, provided there was a large energy input into the ICM from AGN over the entire formation history of haloes. The AGN heating is more efficient at driving X-ray emitting gas from the central regions of lower-mass haloes, reducing their luminosity and steepening the \lxt\ relation as desired. Their model was also able to account for some of the increased scatter about the mean relation seen for temperatures $T\lesssim 3$ keV, attributable to the varied merger histories of groups. 

A limitation of the method of \citet{SHT09} is that radiative cooling is not incorporated in their hydrodynamical simulations. Instead, their model relies on the simplistic treatment of the distribution and cooling of gas employed in existing SAMs. Note, however, that gas particles are still converted to dissipationless `star' particles as dictated by the SAM. While cooling is relatively unimportant for the majority of the ICM, the low central entropy found in CC clusters cannot be reproduced in their simulations, so the large scatter towards the upper-luminosity edge of the observed local \lxt\ relation is not recovered. This will be less of an issue at high-redshift since only a small fraction of clusters have cool cores at $z\gtrsim 0.5$ (e.g. \citealt{VBF07}), due to the higher rate of major mergers (e.g. \citealt{JCB05}).

A more self-consistent approach would be to fully couple a SAM to a radiative simulation, so that the gas distribution in the simulation is used to inform the SAM. This extension of the semi-analytic technique would require the simulation and the SAM to be coupled in such a way that both can be undertaken simultaneously. Extensive testing would be necessary to ensure that such a model was as successful as current SAMs in reproducing observed galaxy properties. Such a scheme is a long-term goal of our work but is beyond the scope of this paper. 

In order to elucidate the effect of feedback from galaxy formation on the evolution of scaling laws, we compare the predictions of our feedback model with those of two other models in the Millennium Gas series. The first of these simulations incorporates gravitational heating only and is thus referred to as the GO run. This is useful as a base model. Given that the only source of gas entropy changes in the GO run is gravity, we would expect a self-similar cluster population to be formed. This is generally found to be the case in such simulations (e.g. \citealt{NFW95,ENF98,VKB05,ASY06,SRE10}). 

The second reference simulation includes high-redshift preheating and radiative cooling, in addition to shock heating. We name this the PC run. Preheating raises the entropy of the ICM before gravitational collapse, preventing gas from reaching high densities in central cluster regions and thus reducing its X-ray emissivity. This effect is greater in lower-mass systems, breaking the self-similarity of the cluster scaling relations in a way that resembles observations \citep{BEM01,BRM01,MTK02,BGW02,TBS03,BFK05,HGP08,SRE10}. However, preheating fails to account for the observed scatter about the mean relations, particularly on group scales, and generates overly-large isentropic cores in low-mass systems (e.g. \citealt{PSF03,PAP06}).

The cosmological model adopted in all three Millennium Gas simulations is a spatially-flat \lcdm\ model with parameters $\Omega_{\rm m,0}=0.25$, $\Omega_{\rm b,0}=0.045$, $\Omega_{\Lambda,0}=0.75$, $h=0.73$, $n_{\rm s}=1$ and $\sigma_{8,0}=0.9$. Here $\Omega_\mathrm{m,0}$, $\Omega_{\rm b,0}$ and $\Omega_{\Lambda,0}$ are the total matter, baryon and dark energy density parameters, respectively, $h$ is the Hubble parameter $H_0$ in units of $100$ km s$^{-1}$ Mpc$^{-1}$, $n_\mathrm{s}$ is the spectral index of primordial density perturbations, and $\sigma_{8,0}$ is the rms linear density fluctuation within a sphere of comoving radius $8h^{-1}$ Mpc. The subscript $0$ signifies the value of a quantity at the present day. These cosmological parameters are the same as those used in the original Millennium simulation and are consistent with a combined analysis of the first-year \emph{Wilkinson Microwave Anisotropy Probe} (WMAP) data \citep{SVP03} and data from the \emph{Two-degree-Field Galaxy Redshift Survey} \citep{CDM01}. However, there is some tension between the chosen parameter values, particularly $n_{\rm s}$ and $\sigma_{8,0}$, and those derived from the seven-year WMAP data \citep{KSD10}. 

We now describe further details of our simulations. The GO and PC runs have already been discussed elsewhere \citep{HGP08,SRE10}, so here we only briefly summarise their properties, focusing our attention mainly on the new FO run.

\subsection{The GO and PC simulations}
\label{sec:GOPC}

Initial conditions for the GO and PC runs were created at a redshift $z_{\rm i}=49$ by displacing particles from a glass-like distribution, so as to form a random realisation of a density field with a \lcdm\ linear power spectrum obtained from CMBFAST \citep{SEZ96}. The amplitudes and phases of the initial perturbations were chosen to match those of the Millennium simulation. In both cases, the simulation volume is a comoving cube of side length $L=500h^{-1}$ Mpc, as in the Millennium Simulation, containing $5\times 10^8$ dark matter particles of mass $m_{\rm DM}=1.42\times 10^{10}h^{-1}\Msun$, and $5\times 10^{8}$ gas particles of mass $m_{\rm gas}=3.12\times 10^{9}h^{-1}\Msun$. These are some of the largest hydrodynamical \nb\ simulations ever carried out. Although the mass resolution is approximately $20$ time coarser than the Millennium Simulation, over $95\%$ of haloes formed with a mass greater than $10^{13}h^{-1}\Msun$ are within $100h^{-1}$ kpc and $5\%$ of their original position and mass, respectively.

The massively parallel TreePM $N$-body/SPH code GADGET-2 \citep{SPR05} was used to evolve both sets of initial conditions to $z=0$, with full particle data stored at $160$ output redshifts. The Plummer-equivalent gravitational softening length was fixed at $\epsilon=100h^{-1}$ kpc in comoving coordinates until $z=3$, then fixed in physical coordinates thereafter. The softening is thus approximately $4\%$ of the mean interparticle spacing, which has been shown to be the optimal choice for hydrodynamical simulations \citep{THC92,BDM06}.

The simple model of preheating employed in the PC simulation is similar to that of \citet{BGW02}. Briefly, the entropy of every particle is raised to $200$ keV cm$^2$ at $z=4$, thus creating an entropy `floor'. In addition to preheating, there is also radiative cooling based on the cooling function of \citet{SUD93}, assuming a fixed metallicity of $0.3\Zsun$ (a good approximation to the mean metallicity of the ICM out to at least $z=1$; \citealt{TRE03}). Once the temperature of a gas particle drops below $2\times 10^4$ K, the hydrogen density exceeds $\rho_{\rm H}=4.2\times 10^{-27}$ g cm$^{-3}$ and the density contrast is greater than $100$, then it is converted to a collisionless star particle. However, the preheating is so extreme that star formation is effectively terminated at $z=4$, so that less than $2\%$ of the baryons are locked-up in stars at $z=0$.
 
\subsection{The FO simulation}
\label{sec:FO}

We have adopted a different approach for the FO simulation than for the other two Millennium Gas runs. Rather than simulating the entire Millennium volume, we decided instead to resimulate a sample of several hundred galaxy groups and clusters extracted from the original Millennium simulation. In this way, it will be less time/resource consuming to develop and test future extensions of the model of \citet{SHT09}. 

There are three distinct components of the FO run: a dark matter-only resimulation of each region containing a cluster from our sample, semi-analytic galaxy catalogues built on the halo merger trees of these resimulations, and hydrodynamical resimulations of the same regions to track the energy injection from model galaxies. We now discuss each stage of the modelling process in turn.

\subsubsection{Dark matter cluster resimulations}

Our cluster sample consists of $337$ objects identified in the $z=0$ output of the Millennium Simulation, spanning a broad mass range: $1.7\times 10^{13}h^{-1}\Msun\leq M_{\rm vir}\leq 2.9\times 10^{15}h^{-1}\Msun$. The sample was constructed as follows. All clusters with $M_{\rm vir}>5\times 10^{14}h^{-1}\Msun$ were selected while, at lower masses, a fixed number of clusters were chosen at random per logarithmic interval in virial mass. For each cluster in the sample, we extract a spherical region from the $z=0$ Millennium snapshot that is centred on the object in question and has a radius equal to twice the cluster virial radius. The particles contained within this sphere are traced back to $z_{\rm i}=127$ (the initial redshift of the Millennium Simulation) to identify the Lagrangian region from which the object formed. Multi-mass initial conditions \citep{TBW97} are then made by following a procedure similar to that of \citet{SWV08}, but with one major difference: we do not increase the mass resolution in the Lagrangian region of interest, but degrade the resolution exterior to this region instead. Particles in the `high-resolution' region then have (approximately) the same mass as in the parent Millennium Simulation, $m_{\rm DM}=8.61\times 10^8h^{-1}\Msun$, but more distant regions are sampled with progressively more massive particles. Our resimulations can thus be thought of as `desimulations'. A glass-like configuration was used for the unperturbed particle distribution in the high-resolution regions of all our initial conditions, and the initial particle displacements were imprinted using the Zeldovich approximation. The amplitudes and phases of the initial perturbations again match those of the Millennium Simulation. 

We adopt the same mass resolution as the Millennium Simulation in the high-resolution regions of our initial conditions for two reasons. First, the properties of our resimulated clusters should then agree well with those of their counterparts in the original Millennium Simulation, except for small perturbations induced by different representations of the tidal field. Second, we will be able to construct more comprehensive galaxy catalogues, and thus a more detailed model for feedback from galaxies, than if we had used the coarser mass resolution of the other Millennium Gas runs.

The initial conditions for each cluster in our sample are then evolved to the present day with GADGET-2. Raw particle data is stored at the $64$ output redshifts of the Millennium Simulation. At each output time, dark matter haloes are identified as virialised groups of high-resolution particles using a parallel friends-of-friends (FOF) algorithm. We adopt a standard FOF linking length of $20\%$ of the mean particle separation \citep{DEF85}, and only save groups that contain at least 20 particles, so that the minimum halo mass is $1.7\times 10^{10}h^{-1}\Msun$. Gravitationally-bound substructures orbiting within these FOF haloes are then found with a parallel version of the SUBFIND algorithm \citep{SWT01}. The resulting catalogues of groups and subgroups are written out alongside the snapshot data.

In the Millennium Simulation, the gravitational force law was softened isotropically on a fixed comoving scale of $\epsilon=5h^{-1}$ kpc (Plummer-equivalent), corresponding to approximately $2\%$ of the mean interparticle spacing. We employ a different softening scheme in our cluster resimulations. The Plummer equivalent gravitational softening length is fixed to $\epsilon=9.2h^{-1}$ kpc in physical coordinates below $z=3$, and to $\epsilon=36.8h^{-1}$ kpc in comoving coordinates at higher redshifts (for the high-resolution particles). The $z=0$ softening length is thus $4\%$ of the mean particle separation. Our choice of a larger softening scale allows us to resolve more low-mass subhaloes, because two-body heating effects are less important, so we can construct more detailed semi-analytic galaxy catalogues.

It is important that the high-resolution region remains free from contamination by more massive boundary particles during the course of a resimulation. Since we are interested in the evolution of cluster properties within $r_{500}$, we have checked whether there are any boundary particles in this region for every cluster used in this study (see Section \ref{sec:samples} below for a description of how we construct cluster samples from our resimulations). Only three objects were found to contain boundary particles interior to $r_{500}$, of which two were significantly contaminated. We discarded these two objects from our cluster samples.

\subsubsection{Feedback from a galaxy formation model}

Dark matter halo merger trees are constructed in post-processing for each resimulated region using the stored subhalo catalogues. This is done by exploiting the fact that each halo will have a unique descendant in a hierarchical scenario of structure formation; see \citet{SWJ05} for details of the procedure. Using these merger trees, we have generated galaxy catalogues for all our resimulations by applying the Munich L-Galaxies SAM of \citet{DLB07}. Galaxy properties are saved at the same $64$ redshifts as the simulation data. We adopted the same set of model parameters as \citet{DLB07}. For a full description of the physical processes incorporated in L-Galaxies, we refer the reader to \citet{CSW06} and \citet{DLB07}.

The information contained within the semi-analytic galaxy catalogues enables us to calculate energy feedback from galaxies into the ICM. We only give an outline the procedure here, since a complete account is given in \citet{SHT09}.

For each cluster in our sample, we first identify all model galaxies that lie within a distance $r_{\rm vir}$ of the centre of the cluster halo at $z=0$, then use the galaxy merger trees to find all their progenitors. We only consider feedback from this subset of galaxies since this is sufficient to correctly determine the properties of the ICM within $r_{500}$, the region we are interested in. For each of these galaxies, we use its merger tree to compute the change in stellar mass, $\Delta M_*$, and mass accreted by the central black hole, $\Delta M_{\rm BH}$, over the time interval $\Delta t$ between successive model outputs. Knowledge of $\Delta M_*$ enables us to incorporate star formation in our simulations as described in the following section. 

We compute the energy imparted to intracluster gas by Type II SNe using the L-Galaxies supernova feedback model. In this model, the total amount of energy released by SNe in a time $\Delta t$ is proportional to $\Delta M_*$. However, some of this energy is assumed to be used up reheating cold gas in the galactic disk. Any energy remaining after reheating is used to eject gas from the halo into the surrounding medium, heating the ICM (see equation 20 of \citealt{CSW06}). 

The model of AGN feedback we employ is that of \citet{BMB08}. In this scheme, the heat energy input into the ICM by AGN over a time period $\Delta t$ is the minimum of
\begin{equation}
\Delta E_{\rm heat}=\epsilon_{\rm r}\Delta M_{\rm BH}c^2
\end{equation}
and
\begin{equation}
\label{eq:AGNcap}
\Delta E_{\rm heat}=\epsilon_{\rm SMBH}\Delta E_{\rm Edd},
\end{equation}
where $c$ is the speed of light, $\Delta E_{\rm Edd}$ is the energy released by a black hole accreting at the Eddington rate in a time $\Delta t$, and $\epsilon_{\rm r}$ is the efficiency with which matter can be converted to energy near the event horizon. Following convention, we set $\epsilon_{\rm r}=0.1$, which is appropriate for radiatively efficient accretion onto a non-rapidly spinning black hole \citep{SHS73}. The parameter $\epsilon_{\rm SMBH}$ is related to the structure of the accretion disk itself. At low accretion rates, the accretion disk is expected to be geometrically thick and advection dominated, enabling efficient jet production and effective radio mode feedback. As the accretion rate is increased, it is thought that the vertical height of the disk eventually collapses, so much more of the accretion energy is radiated away. \citet{BMB08} assume that this change in disk structure occurs once the accretion rate reaches $\dot{M}_{\rm BH}=\epsilon_{\rm SMBH}\dot{M}_{\rm Edd}$, where $\epsilon_{\rm SMBH}=0.02$, leading to an upper limit on the amount of energy available for heating intracluster gas (equation \ref{eq:AGNcap}).

\subsubsection{Hydrodynamical cluster resimulations}

To track the effect of energy feedback from galaxies on the thermodynamical properties of the ICM, we couple the L-Galaxies SAM to hydrodynamical resimulations of our clusters. We use the same multi-mass initial conditions for these resimulations as described above, but we add gas particles with zero gravitational mass. This ensures that the dark matter distribution remains undisturbed by the inclusion of baryons, so that the halo merger trees used to generate the semi-analytic galaxy catalogues will be the same. Although baryons can influence the structure and merger histories of dark haloes \citep{SRE09,SDD08,RSH09,PTS10,DSK10}, such effects are neglected in all modern SAMs, which are founded on merger trees derived from the dark matter distribution. We are forced to follow this route so that we can use SAM input into our simulations. This approximation is unlikely to have any significant impact on the results presented in this paper.

The number of gas particles added to each set of initial conditions is chosen in such a way that their true mass would be the same as in the other two Millennium Gas runs. Note that gas particles are only added to the high-resolution region. We include gas at a lower resolution than the dark matter simply to ease the computational cost of our simulations. The resolution we have adopted is sufficient to obtain numerically-converged estimates of bulk cluster properties for systems with $T\gtrsim 2$ keV \citep{SHT09}.

Each set of initial conditions is evolved from $z_{\rm i}=127$ to $z=0$ with a version of GADGET-2 that has been modified to accommodate gas particles with zero gravitational mass. As in \citet{SHT09}, cooling processes are neglected. The same softening scheme as used in the dark matter resimulations is applied to the gas particles since they do not influence the gravitational dynamics. Whenever an SPH calculation is required, we assign gas particles their true mass, so that gas properties are computed correctly. Gas particles are also given their true mass for simulation data dumps, with the mass of the dark matter particles accordingly reduced to $(1-f_{\rm b})m_{\rm DM}=7.06\times 10^8 h^{-1}\Msun$, where $f_{\rm b}=\Omega_{\rm b,0}/\Omega_{\rm m,0}=0.18$ is the mean cosmic baryon fraction in our cosmological model.

Once an output redshift is reached, temporary `galaxy' particles are introduced throughout the high-resolution region at positions specified by the SAM galaxy catalogue. For each galaxy, we know the increase in stellar mass and energy released by SNe and AGN since the last output (see above). We use this information to form stars and heat gas in the vicinity of model galaxies as detailed in \citet{SHT09}. Briefly, the $\Delta N_{\rm star}=\Delta M_*/m_{\rm gas}$ gas particles nearest to each galaxy are converted into collisionless star particles, where a stochastic method is used to ensure $\Delta N_{\rm star}$ is an integer. Once star formation is complete, the heat energy available from SNe and AGN is distributed amongst all gas particles contained within a sphere of radius $r_{200}$ centred on the galaxy, in such a way that each particle receives an equal entropy boost. If the galaxy is not the central galaxy of a FOF group, then L-Galaxies approximates $r_{200}$ by the radius of a sphere enclosing a mass equal to the product of the number of particles in the host subhalo and the particle mass. In this case, we distribute the available heat energy within this radius instead. The total energy contributed by such galaxies is about an order of magnitude less than for central galaxies anyway. Following the injection of entropy, the galaxy particles are removed and the simulation continues until the next output time, when the process is repeated. Note that increasing the frequency with which energy is injected into ICM has a negligible impact on our results, because the time interval between our chosen $64$ model outputs is always less than the galaxy halo dynamical time.

\subsection{Constructing cluster catalogues and profiles}
\label{sec:samples}

Cluster catalogues are generated at several redshifts for the three Millennium Gas simulations using a procedure similar to that employed by \citet{MTK02}, which we now briefly describe. 

The first step is to identify gravitationally-bound groups of dark matter particles with the FOF algorithm. This was done on the fly in the FO run, and we have group catalogues stored at all $28$ output redshifts between $z=1.5$ and $z=0$ for each resimulated cluster. For the GO and PC runs, FOF groups were identified in post-processing, setting the linking length to be $10\%$ of the mean interparticle separation. Only groups with $500$ particles or more were kept, corresponding to a minimum halo mass of $7.10\times 10^{12}h^{-1}\Msun$. We produced group catalogues for these two simulations at seven different redshifts: $z=1.5$, $1.25$, $1$, $0.75$, $0.5$, $0.25$, $0$. 

The spherical overdensity method is then used to construct cluster catalogues. Briefly, a sphere is grown about the most gravitationally-bound dark matter particle of each FOF group until radii are found that enclose mean overdensities of $\Delta_{\rm vir}(z)$, $\Delta=200$, $\Delta=500$, $\Delta=1000$ and $\Delta=2500$, relative to the critical density $\rho_{\rm cr}(z)$. In cases where clusters overlap, we only keep the object with the largest mass within $r_{2500}$. We also discard clusters with fewer than $1000$ particles at each overdensity, which corresponds to a minimum cluster mass of $8.61\times 10^{11}h^{-1}\Msun$ in the FO run, and $1.73\times 10^{13}h^{-1}\Msun$ in the GO and PC runs.

During the cluster identification process we compute a variety of cluster properties, averaged within each choice of scale radius. The relevant properties for this work are the total mass, gas mass, temperature and X-ray luminosity. The measure of temperature we adopt is the spectroscopic-like temperature \tsl\ \citep{MRM04}. In the Bremsstrahlung regime ($T>2$ keV), this temperature estimator has been shown to provide the closest match to the actual spectroscopic temperature, $T_{\rm spec}$, obtained by fitting X-ray spectra of simulated clusters with a single-temperature plasma model. The X-ray luminosity is approximated by the bolometric emission-weighted luminosity, assuming the cooling function of \citet{SUD93} and a fixed metallicity of $0.3\Zsun$. Three-dimensional gas density, spectroscopic-like temperature and entropy profiles are also computed for all our clusters by averaging particle properties within spherical shells, centred on the minimum of the dark matter potential.

In this paper we focus our attention on the evolution of X-ray scaling relations for massive clusters only, since it is the most massive objects that are observed at high redshift. Previous numerical studies that use a smaller simulation volume than the Millennium volume have been unable to resolve sufficiently large numbers of massive clusters to investigate this in detail. Limiting our scope in this way has two additional benefits. First, the cooling times are longer in the central regions of massive clusters than in groups, so the lack of self-consistent cooling in our feedback model is less of an issue. Second, the X-ray emission will be predominantly thermal Bremsstrahlung, so the spectroscopic-like temperature provides an accurate estimate of the spectral temperature.

We construct samples of massive clusters for our study from the cluster catalogues as follows. The starting point is to discard all clusters whose average density is too low for $r_{2500}$ to be defined. We do this to remove any low-mass objects that may have erroneous properties due to being subclumps falling into more massive neighbouring systems. For scaling laws involving the total cluster mass (\yxm, \tslm\ and \lxm), we then impose a mass cut of $M_{500}\geq 10^{14}h^{-1}\Msun$ at all redshifts of interest. For the \lxtsl\ relation, a cut is made in \tsl\ instead to ensure completeness in \tsl; only clusters with a spectroscopic-like temperature greater than that corresponding to a mass of $M_{500}=10^{14}h^{-1}\Msun$ on the mean \tslm\ relation are kept. Table \ref{tab:numclus} lists the number of clusters in our final samples as a function of redshift for each of the Millennium Gas simulations.

\begin{table}
\caption{Number of clusters in each of the Millennium Gas simulations as a function of redshift, once the mass/temperature cut appropriate for each scaling relation has been made (see text).}
\label{tab:numclus}
\begin{tabular}{@{}lcccc}
\hline
Relation & & Redshift & \\
& 1.5 & 1 & 0.5 & 0 \\
\hline
GO simulation & & & & \\
\yxm, \tslm, \lxm\ & 25 & 145 & 549 & 1109 \\
\lxtsl\ & 15 & 107 & 441 & 946 \\
\\
PC simulation & & & & \\
\yxm, \tslm, \lxm\ & 14 & 102 & 410 & 881 \\
\lxtsl\ & 13 & 93 & 376 & 838 \\
\\
FO simulation & & & & \\
\yxm, \tslm, \lxm\ & 18 & 75 & 148 & 187 \\
\lxtsl\ & 15 & 67 & 139 & 186 \\
\hline
\end{tabular}
\end{table}

\section{Results and discussion}
\label{sec:results}

We now use the samples of massive clusters extracted from our three simulations to investigate differences in evolutionary behaviour that arise from adopting a plausible model for feedback from SNe and AGN, rather than simple preheating or gravitational heating. We first present results at $z=0$ as they will form the basis for measuring the evolution of the thermal properties of the ICM with redshift.   

Throughout the remainder of this work, all uncertainties are quoted at the $68\%$ confidence level.

\subsection{Cluster profiles at $z=0$}
\label{sec:profsz0}

Radial cluster profiles are more sensitive to the precise manner in which non-gravitational cooling and heating processes are implemented in numerical simulations than X-ray scaling laws. Therefore, we start by examining whether our model for feedback from galaxies is able to explain the temperature and entropy profiles of observed low-redshift clusters. The observational dataset we use is REXCESS \citep{BSP07}, a representative sample of $33$ local ($z<0.2$) clusters drawn from the REFLEX catalogue \citep{BSG04}, all of which have been observed with \emph{XMM-Newton}. Temperature profiles for the REXCESS clusters are presented in \citet{APP09}, and entropy profiles in \citet[hereafter PAP10]{PAP10}. We choose to compare with REXCESS for three reasons. First, REXCESS clusters were selected in luminosity only, thus ensuring no morphological bias, in such a way as to sample the X-ray cluster luminosity function in an optimal manner. Second, distances were optimised in REXCESS so that $r_{500}$ falls well within the \emph{XMM-Newton} field-of-view, increasing the precision of measurements at large radii. Third, the same definition of $r_{500}$ is used as in this work.

To facilitate a fair comparison with our simulated data, we only consider observed clusters with a mass $M_{500}\geq 10^{14}h^{-1}\Msun$. We have also rescaled the observational data to account for the fact that \citet{APP09} and PAP10 assumed a slightly different cosmological model in their analysis. 

In the following, it will prove useful to divide the REXCESS sample into CC and non-cool-core (NCC) systems. As in \citet[hereafter PCA09]{PCA09}, clusters are classified as CC systems if they have a central gas density $E(z)^{-2}n_{\rm e}(0)>4.8\times 10^{-2}h^{1/2}{\rm cm}^{-3}$.

\subsubsection{Temperature profiles}

\begin{figure}
\includegraphics[width=85mm]{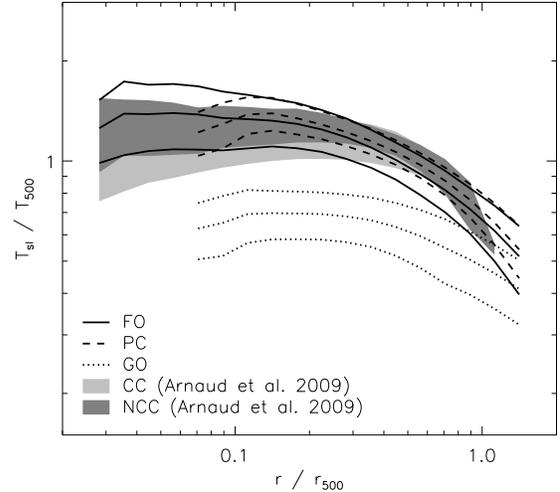}
\caption{Mean spectroscopic-like temperature profiles, with $1\sigma$ scatter, obtained from the Millennium Gas simulations. The light and dark shaded regions enclose the mean profiles, plus $1\sigma$ scatter, of CC and NCC clusters in the REXCESS sample \citep{APP09}, respectively. Only clusters with a mass $M_{500}\geq 10^{14}h^{-1}\Msun$ are considered.}
\label{fig:sltprofz0}
\end{figure}

In Figure \ref{fig:sltprofz0} we display the average spectroscopic-like temperature profile of clusters in the FO simulation. For comparison, we also show average profiles obtained from the reference GO and PC simulations and the observational data of \citet{APP09}. We discard profile data at radii less than the gravitational softening length and only plot the average profile if there are $10$ or more clusters in a given radial bin. All profiles have been normalised to the characteristic halo temperature, $T_{500}$, computed from the self-similar model:
\begin{equation}
\label{eq:T500}
T_{500}=\frac{G}{2}\frac{\mu m_{\rm H}}{k_{\rm B}}\frac{M_{500}}{r_{500}},
\end{equation}
where $\mu =0.59$ is the mean molecular weight for a fully ionised gas of primordial composition and $m_{\rm H}$ is the mass of a hydrogen atom. In the case of the observed profiles, $T_{500}$ is calculated at the redshift of each individual cluster. With this scaling we would expect cluster profiles to coincide in the pure gravitational heating scenario.

Clusters formed in the GO simulation are clearly cooler than observed. This is because the spectroscopic-like temperature estimate is biased low by the cool, low-entropy cores of accreted subhaloes that are prevalent in GO cluster haloes (e.g. \citealt{MAE01}). The temperature profiles of individual clusters are flat in core regions, but we see a slight decline in the average scaled profile at small radii, $r\lesssim 0.1r_{500}$. This is because we are only averaging over the profiles of the most massive clusters at such radii, since the gravitational softening length is a smaller fraction of $r_{500}$ for these objects, and the normalisation of the scaled temperature profiles decreases with increasing mass. In the self-similar model the normalisation of the scaled profiles should not depend on mass; the reason for the mass-dependence in our GO simulation is that more massive clusters are less concentrated than their low-mass counterparts, i.e. even the dark matter is not truly self-similar.

In the PC and FO runs the source of non-gravitational heating is completely different, but the net effect of the entropy injection is the same: cool subclumps are erased and the average temperature of the intracluster gas increases. Both simulations lead to temperature profiles that provide a good overall match to the observed profiles of NCC clusters, being nearly isothermal in core regions $r\lesssim 0.15r_{500}$. Again, the down-turn in the average scaled profiles visible in core regions arises because, for small values of $r/r_{500}$, we are taking an average of the scaled temperature profiles of the most massive objects only, which have a lower normalisation than those of less massive systems.

Recall that we have chosen to neglect gas cooling processes in our FO simulations, so it is not surprising that we do not reproduce the gradual decline in temperature seen in the central parts of CC clusters, where the cooling time is short. Although the PC run does incorporate radiative cooling, it has a negligible effect on the gas temperature since the entropy of the ICM is raised before structure formation commences, preventing gas from reaching high densities in cluster cores and cooling efficiently.

Fully self-consistent simulations with radiative cooling and stellar feedback typically predict temperature profiles with a sharp spike at small cluster-centric radii, followed by a rapid drop in temperature moving further into the core (e.g. \citealt{BMS04,NKV07,SSD07}). This is due to the adiabatic compression of gas flowing in from cluster outskirts to compensate for the lack of pressure support caused by too much gas cooling out of the hot phase. Temperature profiles of this form clearly conflict with the smoothly declining (flat) profiles of observed CC (NCC) clusters (e.g. \citealt{SPO06,VKF06,APP09}). 

Even if a feedback mechanism is able to reproduce observed temperature profiles, this does not guarantee that radiative cooling has been balanced. For example, the stellar feedback scheme employed by \cite{KDA07} is capable of producing temperature profiles that are in reasonable agreement with observational data. However, the resulting stellar fraction within clusters was found to be $\sim 25\%$, far in excess of the observed value of $\sim 10\%$ \citep{BPB01,LMS03,BMBE08}. 

In recent work, \citet{FBT10} have included a sub-grid model for AGN feedback in hydrodynamical simulations. They found that the additional heating from AGN was insufficient to prevent overcooling in massive clusters, again leading to too high a star formation efficiency and sharply peaked temperature profiles. However, on the scale of galaxy groups, their AGN feedback scheme was able to successfully regulate the thermal structure of the ICM; see also \citet{MSP10}.

\subsubsection{Entropy profiles}

The entropy of intracluster gas increases when heat energy is introduced, and decreases when radiative cooling carries heat energy away. Entropy profiles thus preserve a record of the physical processes responsible for similarity breaking in clusters (e.g. \citealt{VBB02,VBB03}). 

If shock heating were the only mechanism acting to raise the entropy of the gas, then analytical models based on spherical collapse predict that entropy scales with radius as $K\propto r^{1.1}$ outside of central cluster regions \citep{TON01}. Cosmological simulations that only include gravitational heating give rise to slightly steeper entropy profiles in cluster outskirts: $K\propto r^{1.2}$ (e.g. \citealt{VKB05, NKV07}). 

Observed profiles are also typically found to scale as $K\propto r^{1.1}$ at large cluster-centric radii, flattening in central regions (e.g. \citealt{PSF03,SVD09,CDV09,SOP09}, PAP10). However, the precise radius at which this flattening occurs varies considerably, depending on such factors as the temperature (mass) of the system and whether it has a CC or a NCC. In particular, hotter, more massive objects have a higher mean core entropy (e.g. \citealt{CDV09}), and the profiles of NCC clusters flatten off at significantly larger radii than those of CC clusters  (e.g. \citealt{SOP09}, PAP10).

\begin{figure}
\includegraphics[width=85mm]{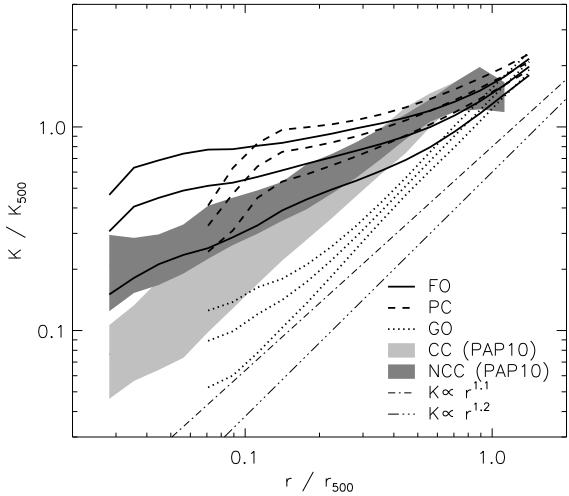}
\caption{Mean entropy profiles, with $1\sigma$ scatter, obtained from the Millennium Gas simulations. The light and dark shaded regions enclose the mean profiles, plus $1\sigma$ scatter, of observed CC and NCC clusters from REXCESS (PAP10), respectively. We only consider clusters with a mass $M_{500}\geq 10^{14}h^{-1}\Msun$.}
\label{fig:Kprofz0}
\end{figure}

Figure \ref{fig:Kprofz0} compares the entropy profiles obtained from each of our three simulations with observational data from REXCESS (PAP10). For illustrative purposes, we also show the power-laws $K\propto r^{1.1}$ and $K\propto r^{1.2}$, assuming an arbitrary normalisation in both cases. We have scaled all entropy profiles by the `virial' entropy, $K_{500}$, defined as
\begin{equation}
\label{eq:K500}
K_{500}=\frac{k_{\rm B}T_{500}}{n_{{\rm e},500}\, ^{\gamma-1}},
\end{equation}
where $n_{{\rm e},500}$ is the average electron density within $r_{500}$, given by
\begin{equation}
\label{eq:ne500}
n_{\rm e,500}=\frac{500f_{\rm b}E(z)^2\rho_{\rm cr,0}}{\mu_{\rm e}m_{\rm H}},
\end{equation}
and $\mu_{\rm e}=1.14$ is the mean molecular weight per free electron. Note that $K_{500}$ depends only on the total halo mass, so is independent of the thermodynamic state of the gas. 

The entropy profiles of clusters extracted from the GO run are indeed well described by the power-law $K\propto r^{1.2}$ for $r\gtrsim 0.15r_{500}$, agreeing with the results of previous studies. There is also very little scatter about the mean, indicating self-similar scaling. For radii interior to $0.15r_{500}$ there is more diversity; some of the clusters have nearly isentropic cores while others show no signs of flattening. Compared to the observed entropy profiles of CC clusters, the profiles of objects in the GO run have a similar slope, at least for $r\gtrsim 0.15r_{500}$, but the normalisation is systematically too low. In the case of NCC clusters, it is evident that the GO model cannot explain the shallow profiles characteristic of these systems. 

Clusters formed in the PC and FO simulations have entropy profiles that are broadly consistent with the theoretical scaling $K\propto r^{1.1}$ at large radii $r\gtrsim r_{500}$. This supports the idea that gravity dominates the ICM thermodynamics in the outer regions of clusters. As we move in towards the core from $r_{500}$, the slope decreases and the profiles flatten off, providing a fair match to the observed entropy profiles of NCC clusters. However, on average, both the PC and FO models predict entropy profiles with a shallower slope than those of NCC clusters in central regions, resulting in an overestimate of the core entropy. Note that the FO run yields entropy profiles that are slightly closer to the observed NCC cluster profiles than the PC run. 

\begin{figure}
\includegraphics[width=85mm]{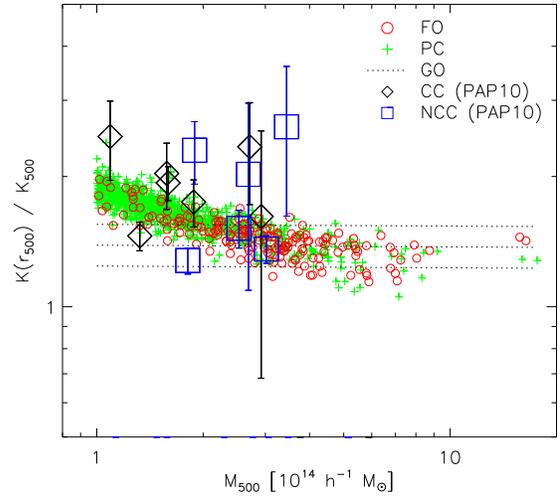}
\caption{Scaled entropy at $r_{500}$ as a function of total mass within $r_{500}$ for clusters in the Millennium Gas simulations. For clarity, we do not display individual data points for the GO run, but instead show the best-fit relation and the typical dispersion about this relation. Observational data for CC and NCC clusters in the REXCESS sample (PAP10) are also shown for comparison, along with $1\sigma$ error bars. Only clusters with $M_{500}\geq 10^{14}h^{-1}\Msun$ are considered. }
\label{fig:Kscalerel}
\end{figure}

In both the PC and FO runs, we see a drop in the average scaled entropy profiles at small values of $r/r_{500}$, where we are averaging over the profiles of just the most massive systems. This is because the normalisation of cluster entropy profiles decreases with mass in these simulations. To demonstrate this, in Figure \ref{fig:Kscalerel} we plot the scaled entropy at $r_{500}$ as a function of $M_{500}$ for each of our simulations, along with observational data from REXCESS (PAP10). For clarity, we do not plot individual points for the GO simulation, but instead show the best-fit relation in log-log space, and the typical dispersion about this relation. The gradient of the best-fit line is very close to zero, so the normalisation of the scaled entropy profiles is independent of mass for GO clusters, as in the self-similar model. By contrast, our PC and FO models predict that the scaled entropy at $r_{500}$ is a decreasing function of mass, implying that non-gravitational heating affects the entropy structure of the ICM out to larger radii in lower-mass systems. This is consistent with the expectation that non-gravitational processes are more influential at the low-mass end of the cluster population. Massive clusters in the PC and FO runs have a similar entropy at $r_{500}$ to their GO counterparts. Note that the observational data points also appear to suggest that the scaled entropy at $r_{500}$ decreases with increasing mass, being scattered about the PC and FO model predictions.

The mass-dependence of the normalisation of scaled cluster entropy profiles in the PC and FO runs explains why we see a larger scatter about the average profile than in the GO run. The scatter about the mean profile in the PC run is similar to that found in observed profiles of NCC systems, but the FO run generates clusters with a wider range of entropy profiles, leading to a larger scatter than is observed.

Neither the PC or the FO models are capable of reproducing the steeply declining entropy profiles seen in CC clusters. In the case of the FO run, this problem could potentially be overcome by including radiative cooling in our model, since cooling acts to lower the entropy in dense central regions where the gas cooling time is short. As we have said, cooling is included in the PC run, but it is curtailed at high redshift by the preheating.

Self-consistent $N$-body/SPH simulations that incorporate cooling, star formation and associated feedback are able to produce entropy profiles that resemble those of CC clusters, with a normalisation in the outer parts of clusters that is higher than predicted by pure gravitational heating, and a steep slope that remains roughly constant all the way into the core (e.g. \citealt{BGW02,BMS04,KTJ04,KAY04,KDA07}). However, this success is usually achieved at the expense of excessive star formation. Furthermore, such simulations fail to reproduce the observed entropy profiles of NCC systems. 

\subsection{X-ray scaling relations at $z=0$}

We now discuss whether our feedback model generates local X-ray scaling laws that are compatible with observations, focusing on the \yxm, \tslm, \lxm\ and \lxtsl\ relations. In each case, we compare with the corresponding relation derived from the low-redshift REXCESS data by PCA09. Wherever possible, we explain any differences that arise using the knowledge gleaned from our discussion of cluster profiles. The data of PCA09 are particularly suitable for a comparison with our simulated cluster samples because they  tabulate spectral temperatures and luminosities within $r_{500}$ (where $r_{500}$ is defined as in this work), and their luminosities are bolometric. Note that we have rescaled the observational data to allow for the slightly different choice of cosmological parameters adopted by PCA09. 

For each set of cluster properties, $(X,Y)=(M,\yx)$, $(M,\tsl)$, $(M,\lx)$ and $(\tsl,\lx)$, we fit a power-law scaling relation of the form 
\begin{equation}
\label{eq:genscalerel}
E(z)^n Y=C_0\left(\frac{X}{X_0}\right)^{\alpha},
\end{equation}
to our simulated data points by minimising $\chi^2$ in log space. Here $X_0=5\times 10^{14}h^{-1}\Msun$ if $X=M$ and $X_0=6\ {\rm keV}$ if $X=\tsl$. The normalisation $C_0$ has units of $10^{14}\Msun \ {\rm keV}$, ${\rm keV}$ and $10^{44}h^{-2}\ {\rm erg\ s}^{-1}$ for $Y=\yx$, $\tsl$ and $\lx$, respectively. Best-fitting parameters $\alpha$ and $C_0$ for each relation are summarised in Table \ref{tab:scalerelz0}. The factor $E(z)^n$ is included to remove the self-similar evolution predicted by equations (\ref{eq:TM})--(\ref{eq:YM}), where the index $n=-2/3$, $-2/3$, $-7/3$ and $-1$ for the \yxm, \tslm, \lxm\ and \lxtsl\ relations, respectively. We include this scaling factor simply to `adjust' observational data to $z=0$ for comparison with our $z=0$ simulated clusters. This will only be a small effect for the redshift range ($z<0.2$) probed by the REXCESS sample.

Scatter in the relations, $\sigma_{\log_{10}Y}$, is quanitfied via the rms deviation of $\log_{10}{Y}$ from the mean relation:

\begin{equation}
\label{eq:sig}
\sigma_{\log_{10}{Y}}^2=\frac{1}{N-2}\sum_{i=1}^{N}\left[\log_{10}{Y_i}-\alpha\log_{10}{\left(\frac{X}{X_0}\right)-\log_{10}{C_0}}\right]^2,
\end{equation} 

\n where $N$ is the number of individual data points $(X_i,Y_i)$. The scatter about each relation is also listed in Table \ref{tab:scalerelz0}.

\begin{table}
\caption{Best-fit parameters (with $1\sigma$ errors) for the $z=0$ X-ray scaling relations obtained from the Millennium Gas simulations.}
\label{tab:scalerelz0}
\begin{tabular}{@{}lccc}
\hline
Relation & $C_0$ & $\alpha$ & $\sigma_{\log_{10}Y}$ \\
\hline
GO simulation & & & \\
\yxm & $4.202\pm 0.070$ & $1.547\pm 0.014$ & $0.087$\\
\tslm & $3.931\pm 0.057$ & $0.554\pm 0.012$ & $0.076$\\
\lxm & $18.58\pm 0.53$ & $1.203\pm 0.024$ & $0.148$\\
\lxtsl & $37.6\pm 1.5$ & $2.004\pm 0.040$ & $0.137$\\
\\
PC simulation & & & \\
\yxm & $5.622\pm 0.052$ & $1.7805\pm 0.0079$ & $0.045$\\
\tslm & $6.310\pm 0.031$ & $0.5512\pm 0.0042$ & $0.024$\\
\lxm & $5.549\pm 0.088$ & $1.842\pm 0.013$ & $0.076$\\
\lxtsl & $4.563\pm 0.055$ & $3.297\pm 0.020$ & $0.063$\\
\\
FO simulation & & & \\
\yxm & $5.757\pm 0.069$ & $1.692\pm 0.016$ & $0.048$\\
\tslm & $6.333\pm 0.049$ & $0.521\pm 0.010$ & $0.031$\\
\lxm & $6.17\pm 0.15$ & $1.777\pm 0.033$ & $0.098$\\
\lxtsl & $4.99\pm 0.12$ & $3.296\pm 0.065$ & $0.104$\\
\hline
\end{tabular}

\medskip
$C_0$ is the best-fitting normalisation of the relations, and $\alpha$ is the best-fitting slope; see equation (\ref{eq:genscalerel}). $\sigma_{\log_{10}Y}$ is the scatter about the mean relation as defined by equation (\ref{eq:sig}).
\end{table}

\subsubsection{The \yxm\ relation}

\begin{figure}
\includegraphics[width=85mm]{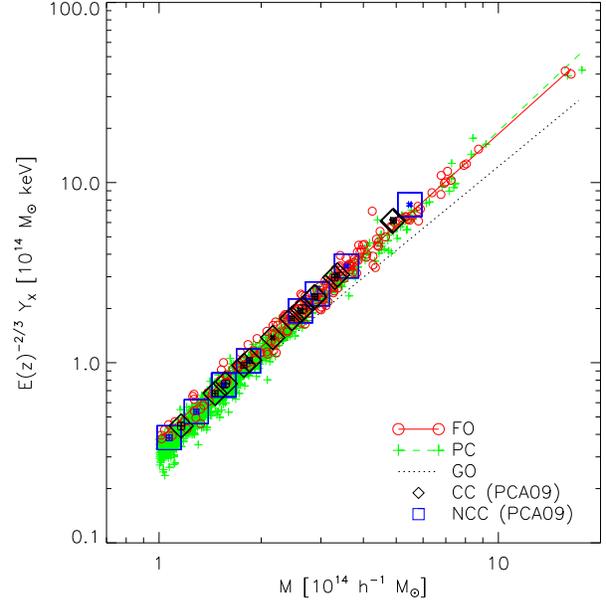}
\caption{\yx\ as a function of total mass within $r_{500}$ for $z=0$ clusters in the Millennium Gas simulations. We do not display data points from the GO run for clarity. Observational data for CC and NCC clusters from REXCESS (PCA09) is shown for comparison, along with $1\sigma$ error bars.}
\label{fig:YM_core}
\end{figure}

The \yxm\ relation is particularly important as both simulations and observations indicate that \yx\ is a low-scatter mass proxy, even in the presence of significant dynamical activity. Figure \ref{fig:YM_core} shows the local \yxm\ relation obtained from the the Millennium Gas FO simulation, along with the relations derived from the GO and PC simulations and the observational data of PCA09. We define \yx\ as the product of the gas mass inside $r_{500}$ and the spectroscopic-like temperature in the $0.15r_{500}<r\leq r_{500}$ region, for consistency with the observations.

In the self-similar model we would expect the slope of the \yxm\ relation to be $\alpha=5/3$ (equation \ref{eq:YM}). The \yxm\ relation obtained from the GO simulation is significantly shallower than both this and the observed relation. The slope of the PC \yxm\ relation is consistent with the observed slope $\alpha\approx 1.8$ from \citet{APP07} at the $1\sigma$ level. On the other hand, the FO run yields a \yxm\ relation that is shallower than observations suggest, with a slope closer to that expected from self-similar scaling. This is similar to the result $\alpha\approx 1.7$ obtained by \citet{KVN06} using simulations with radiative cooling, star formation and supernova feedback. The predicted scatter about the mean relation is similar in the PC and FO runs, being about a factor of $2$ less than in the GO simulation. We note that \yx\ does not appear to be as tightly correlated with cluster mass as \tsl\ in any of our simulations.

The fact that both the PC and FO \yxm\ relations lie close to the self-similar prediction implies that \yx\ must be relatively unaffected by the non-gravitational heating in these models. In the case of the FO run, this is presumably because AGN feedback removes gas from the central regions of haloes, reducing the gas mass within $r_{500}$, but this is offset by an increase in gas temperature caused by the continual injection of entropy from galaxy formation. Preheating evacuates haloes and increases the temperature of the intracluster gas at high redshift instead, but the eventual outcome is essentially the same.

It is important to note that the masses of the REXCESS clusters were not calculated from gas temperature and density profiles using the equation of hydrostatic equilibrium. This is because, by construction, REXCESS contains objects with a wide variety of dynamical states, so hydrostatic equilibrium may be a poor approximation in some cases. PCA09 estimate cluster masses using the \yxm\ relation of \citet{APP07} instead, so the observational data points shown in Figure \ref{fig:YM_core} all lie on this relation by construction.

The \yxm\ relation of \citet{APP07} was calibrated using clusters with hydrostatic mass estimates. Simulations have shown that the assumption of hydrostatic equilibrium can bias such mass estimates low by $\sim 10-20\%$ \citep{REM06,KDA07,NKV07,BHG08,PIV08,MRM09}. This is primarily because of additional pressure support provided by subsonic bulk motions in the ICM. Further bias may be introduced if there is significant pressure support from non-thermal components such as cosmic rays and magnetic fields \citep{LDK10}. Therefore, the masses of clusters in the REXCESS sample are also likely to be underestimated by the same amount.

If the total mass is indeed biased low by $\sim 10-20\%$, then $r_{500}$ will be underestimated by $\sim 3-7\%$. Consequently, the gas mass within $r_{500}$ will be biased low, because it is obtained by integrating the density profile out to $r_{500}$, and the temperature will be overestimated (Figure \ref{fig:sltprofz0}). The former effect will dominate the latter, so we also expect \yx\ to be underestimated. However, this is unlikely to fully compensate for the bias in the total mass estimate, so there may actually be a small offset between our PC and FO \yxm\ relations and the observed relation once any hydrostatic mass bias is accounted for.

\subsubsection{The \tslm\ relation}

\begin{figure}
\includegraphics[width=85mm]{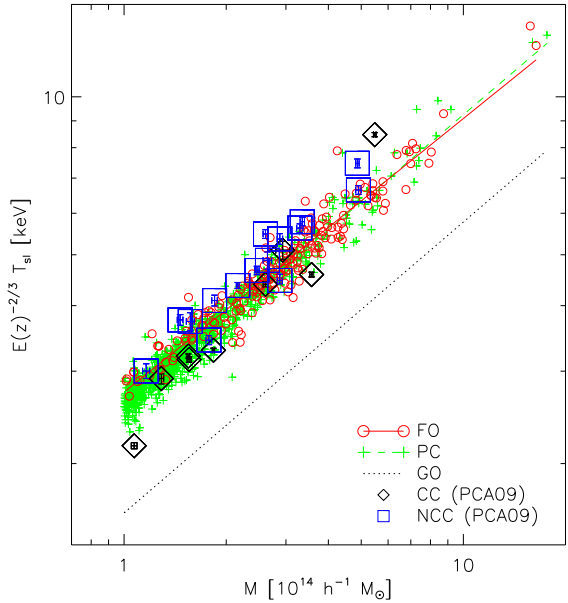}
\caption{Spectroscopic-like temperature as a function of total mass within $r_{500}$ for $z=0$ clusters in the Millennium Gas simulations. We also display observational data for CC and NCC clusters in the REXCESS sample (PCA09).}
\label{fig:TM_core}
\end{figure}

The \tslm\ relations obtained from the GO, PC and FO runs at $z=0$ are compared with observational data from REXCESS in Figure \ref{fig:TM_core}. 
 
For a given mass, clusters in the GO run are much cooler than observed, because the spectroscopic-like temperature estimate is dominated by the cool, dense cores of merging subhaloes. Differences in the distribution of this substructure drive fluctuations in \tsl, leading to a large scatter about the mean relation. The scatter is $\sim 2.5$ ($3$) times larger than in the FO (PC) run. The slope of the GO \tslm\ relation is shallower than observed, and also significantly shallower than $\alpha=2/3$ expected fom the self-similar model (see equation \ref{eq:TM}). The reason for this is that concentration depends on cluster mass in our GO simulation: low-mass clusters are more concentrated, and thus hotter, than their high-mass counterparts, flattening the \tslm\ relation relative to the self-similar prediction.

Non-gravitational heating in the PC and FO runs raises the mean temperature of the ICM above that expected from gravitational heating alone (Figure \ref{fig:sltprofz0}). At a fixed mass, PC and FO clusters are thus hotter than their GO counterparts, with temperatures close to those of observed clusters. Both the PC and FO \tslm\ relations have a similar normalisation and slope, even though the manner in which entropy is injected into the intracluster gas is completely different in each case. The slope $\alpha\approx 0.55$ is close to that derived by \citet{KDA07} from a simulation incorporating gas cooling and stellar feedback. However, the REXCESS relation has a significantly steeper slope: $\alpha=0.633\pm 0.032$ or $\alpha=0.622\pm 0.031$ (G.~W.~Pratt, priv. comm.), depending on the fitting procedure adopted (see PCA09 for details). Note that the observed slope is consistent with the self-similar value, indicating that the $T_{\rm spec}$-$M$ relation is relatively insensitive to baryonic physics. 

Cooling processes are neglected in our FO model and inefficient in the PC model, so systems with a CC are not formed in either simulation. If we only consider the NCC clusters in REXCESS, the resulting $T_{\rm spec}$-$M$ relation is shallower: $\alpha=0.613\pm 0.022$ or $\alpha=0.617\pm 0.022$ (G.~W.~Pratt, priv. comm.), which is slightly closer to, but still steeper than, the slope obtained from our PC and FO simulations. The intrinsic scatter (i.e. once measurement errors have been accounted for) about the REXCESS $T_{\rm spec}$-$M$ relation for NCC clusters is $\sigma_{\log_{10}Y}=0.025\pm 0.015$ (G.~W.~Pratt, priv. comm.), regardless of the regression method used, which is consistent with the scatter about the mean relation obtained from our PC and FO simulations. However, we note that the observational scatter will be an underestimate of the true dispersion since the masses of REXCESS clusters were derived from the \yxm\ relation of \citet{APP07} assuming no intrinsic scatter about that relation.

It appears that a better fit to the observed data could be obtained, at least for the NCC clusters, if the the observational data points in Figure \ref{fig:TM_core} were shifted to the right by $\sim 10\%$ in mass. This is consistent with the level of bias expected from hydrostatic estimates of cluster mass.

\subsubsection{The \lxm\ relation}

\begin{figure}
\includegraphics[width=85mm]{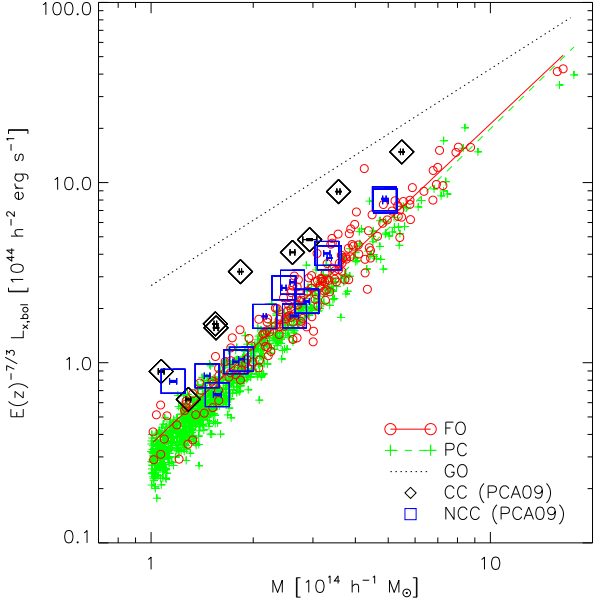}
\caption{Bolometric X-ray luminosity as a function of total mass within $r_{500}$ for $z=0$ clusters in the Millennium Gas simulations. Observational data for CC and NCC clusters from REXCESS (PCA09) is also shown.}
\label{fig:LM_core}
\end{figure}

In Figure \ref{fig:LM_core} we show the three local Millennium Gas \lxm\ scaling relations, plus observational data from PCA09. 

The peak of the X-ray emission in GO clusters is unresolved in our simulation, so the computed luminosities are not trustworthy. We present the GO \lxm\ relation merely to illustrate how dramatically the model fails in reproducing the observational data. 

The \lxm\ relations obtained from the PC and FO simulations are both steeper than anticipated from pure gravitational heating. Both relations have a slope $\alpha\approx 1.8$, whereas the slope measured by PCA09 is $\alpha=1.81\pm 0.10$ or $\alpha=1.96\pm 0.11$. In the case of the FO run, this departure from self-similar scaling arises because AGN feedback expels gas from central cluster regions, reducing the gas density and thus X-ray emissivity. This effect is stronger in less massive systems, steepening the \lxm\ relation as observed. By contrast, similarity-breaking is accomplished in the PC model by boosting the entropy of the ICM before gravitational collapse commences, preventing gas from reaching high densities in cluster cores and lowering the X-ray luminosity. Nevertheless, both the PC and FO models yield almost identical \lxm\ relations at $z=0$. 

The large scatter towards the upper edge of the observed \lxm\ relation is due to systems with a highly X-ray luminous CC. The PC and FO simulations cannot account for this scatter since neither model can reproduce the steeply declining entropy profiles of CC clusters (Figure \ref{fig:Kprofz0}). Quantitatively, PCA09 measure the scatter about the mean \lxm\ relation to be $\sigma_{\log_{10}(Y)}=0.166\pm 0.026$, which is about a factor of $2$ greater than the dispersion about our mean PC and FO \lxm\ relations, even without accounting for the intrinsic scatter about the \yxm\ relation underpinning the REXCESS cluster mass estimates.

Removing the CC clusters from the REXCESS sample leads to a shallower slope, $\alpha=1.705\pm 0.094$ or $\alpha=1.766\pm 0.093$, and a reduction in the measured scatter, $\sigma_{\log_{10}(Y)}=0.094\pm 0.017$, both of which agree very well with the predictions of our FO model. By contrast the PC \lxm\ relation appears to be slightly too steep with too little scatter. The difference in scatter predicted by our two non-gravitational heating models is attributable to the greater diversity in the entropy profiles of objects formed in the FO simulation (Figure \ref{fig:Kprofz0}).  

There is an apparent offset between our PC and FO \lxm\ relations and the observational data for NCC clusters. The magnitude of this offset is about $\sim 10\%$ in mass, which could be accounted for by bias in the observational mass estimates due to the assumption of hydrostatic equilibrium.

\subsubsection{The \lxtsl\ relation}

\begin{figure}
\includegraphics[width=85mm]{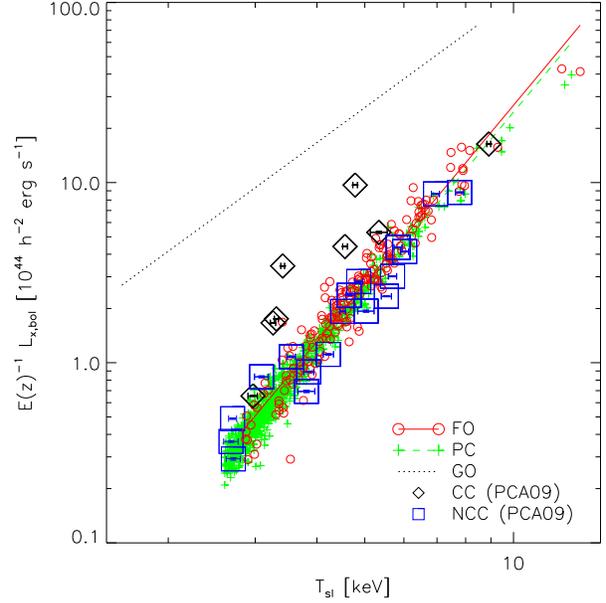}
\caption{Bolometric X-ray luminosity as a function of spectroscopic-like temperature for $z=0$ clusters in the Millennium Gas simulations. X-ray properties are calculated within $r_{500}$. For comparative purposes, we plot observational data for CC and NCC clusters in the REXCESS sample (PCA09).}
\label{fig:LT_core}
\end{figure}

The local GO, PC and FO \lxtsl\ scaling relations are displayed in Figure \ref{fig:LT_core}, along with observational data from REXCESS. Again, the GO relation is only shown for illustrative purposes. 

The PC and FO simulations produce almost identical \lxtsl\ relations, with a steep slope $\alpha\approx 3.3$ owing to the the breaking of self-similarity induced by non-gravitational heating. \citet{KDA07} found a similar slope using a simulation with a self-consistent stellar feedback scheme. For comparison, PCA09 find $\alpha=2.70\pm0.24$ or $\alpha=3.35\pm 0.32$ for the REXCESS sample.

The dispersion about the REXCESS $L_{\rm X}$-$T_{\rm spec}$ relation is $\sigma_{\log_{10}(Y)}=0.288\pm 0.050$ or $\sigma_{\log_{10}(Y)}=0.318\pm 0.059$. This is roughly $3$ ($5$) times larger than the scatter about the FO (PC) \lxtsl\ relation, because no X-ray cores are formed in either simulation. However, the PC and FO \lxtsl\ relations seem to provide a good fit to the NCC clusters in REXCESS. The observed relation for NCC clusters has a slope $\alpha=2.89\pm 0.21$ or $\alpha=3.06\pm 0.19$, with corresponding scatter $\sigma_{\log_{10}(Y)}=0.116\pm 0.025$ or $\sigma_{\log_{10}(Y)}=0.124\pm 0.030$. This is consistent with the scatter about our FO relation, but the PC relation is too tight, with a dispersion that is about of a factor of $2$ less.

\subsection{Evolution of cluster profiles from $z=1.5$ to $z=0$}

We have demonstrated that our feedback model can reproduce the observed properties of massive low-redshift clusters reasonably well, apart from those with a CC. We have also seen that the $z=0$ properties of clusters formed in the feedback run can be replicated almost exactly with a simplistic preheating model, where the entropy of the ICM is raised impulsively at $z=4$, rather than by continual heating from SNe and AGN. Consequently, we cannot discriminate between these two models using data from local observations. 

We now investigate whether feedback from galaxy formation leads to significantly different evolutionary behaviour than simple preheating. In this way we may be able to break the low-redshift degeneracy of the two models. We begin by examining the evolution of cluster profiles from $z=1.5$ to $z=0$; this will help us to understand the predicted evolution of X-ray scaling relations discussed in Section \ref{sec:scalerelevln} below. No attempt is made to compare our results with observations since the available data is, as yet, very limited at high redshift. 

\subsubsection{Results for the PC simulation}

\begin{figure*}
\includegraphics[width=160mm]{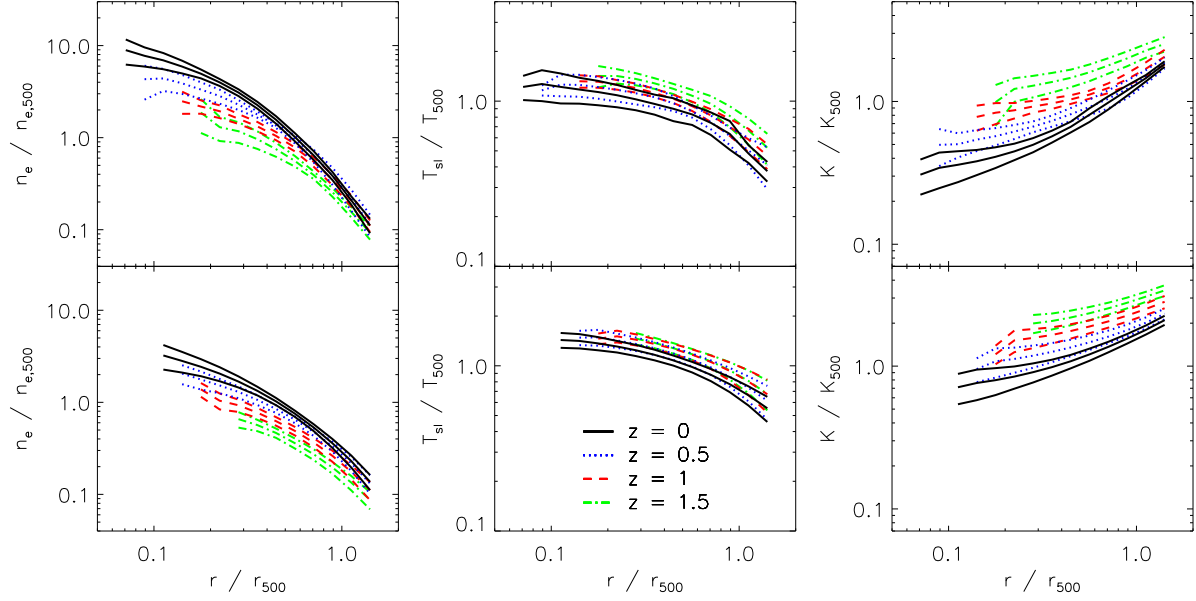}
\caption{Evolution of gas density (first column), spectroscopic-like temperature (second column) and entropy (third column) profiles for clusters in the Millennium Gas PC run. The top row shows the evolution of the ten hottest systems, while the bottom row shows the evolution of clusters with a temperature in the range $3\ {\rm keV}\leq T_{\rm sl}\leq 4\ {\rm keV}$.}
\label{fig:PCprofevln}
\end{figure*}

In Figure \ref{fig:PCprofevln} we show the evolution of gas density, spectroscopic-like temperature and entropy profiles for clusters in the PC run. In the top row, we plot the profiles of the $10$ hottest systems (i.e. the most massive objects) at each redshift and, in the bottom row, profiles for clusters in a narrow temperature range $3\ {\rm keV}\leq T_{\rm sl}\leq 4\ {\rm keV}$. Again, we only keep profile data at radii greater than the gravitational softening length. We scale the density, temperature and entropy profiles by $n_{\rm e,500}$, $T_{500}$ and $K_{500}$, respectively; see equations (\ref{eq:T500})--(\ref{eq:ne500}). With this scaling we would expect to see no evolution of cluster profiles in the self-similar model, where the ICM is only heated by gravitational processes. We have confirmed that this is indeed the case in our GO run.

Focusing on the profiles of the hottest clusters, we see clear signs of evolution beyond the self-similar prediction, which can be understood as follows. Imposing a uniform entropy floor at $z=4$ boosts the entropy of the ICM more in core regions than at large radii. Gas is driven out from central cluster regions, flattening the density profile and increasing the normalisation of the temperature profile, relative to the prediction from gravitational heating alone. Note that the temperature must increase if the slope of the density profile decreases to maintain pressure support. 

After preheating, ejected gas is gradually reincorporated into descendant haloes as hierarchical growth proceeds. Since the gas density has already been lowered by the preheating, it does not decrease as rapidly as in the gravitational heating scenario, so we see an increase relative to the average gas density $n_{\rm e,500}$ as $z\rightarrow 0$. Likewise, the preheated gas has a higher temperature and entropy than if the only source of heating was gravity. Therefore, as gas is accreted back onto descendant haloes, compression and shock heating raise its temperature and entropy at a lesser rate than in the gravitational heating model. This explains why we see a drop in gas temperature and entropy with redshift relative to $T_{500}$ and $K_{500}$, respectively. Given that there is no further non-gravitational heating of intracluster gas in the PC run, the high-redshift entropy injection will become increasingly `diluted' with time, and cluster profiles will eventually resemble those obtained from a simulation with gravitational heating only. 

At any given redshift, several of the $10$ hottest objects are likely to have undergone a recent major merger, which could potentially affect the shape, dispersion and evolution of cluster profiles. To investigate this, we first compute the substructure statistic
\begin{equation}
S=\frac{|\mathbf{x}_{\rm com}-\mathbf{x}_{\rm c}|}{r_{500}},
\end{equation}
for each of the $10$ most massive clusters at each redshift of interest. Here, $\mathbf{x}_{\rm c}$ is the location of the dark matter potential minimum, which we take to be the cluster centre, and $\mathbf{x}_{\rm com}$ is the centre of mass of the cluster gas, defined by
\begin{equation}
\mathbf{x}_{\rm com}=\mathbf{x}_{\rm c}+\frac{\sum_{i}m_{i}(\mathbf{x}_{i}-\mathbf{x}_{\rm c})}{\sum_{i}m_{i}},
\end{equation}
where the sums are over all gas particles within $r_{500}$. Systems undergoing a major merger will be dynamically disturbed and will thus have a larger value of $S$. Following \citet{KDA07}, we say that a cluster is disturbed if $S>0.1$, and relaxed otherwise. At each redshift, we have found that the shape and dispersion of the radial profiles shown in the top row of Figure \ref{fig:PCprofevln} remain almost unchanged if we only consider relaxed clusters in the sample of the $10$ hottest systems. This signifies that our results are not affected by cluster mergers.

The scaled profiles of clusters with $3\ {\rm keV}\leq T_{\rm sl}\leq 4\ {\rm keV}$ evolve in similar way to those of the most massive objects, but their shape is different at each redshift. In particular, their density and entropy profiles are flatter. The reason for these differences is that we are now considering lower-mass clusters. Consequently, preheating is more effective at removing gas from their shallower potential wells, modifying their thermodynamic properties out to larger radii. 

\subsubsection{Results for the FO simulation}

\begin{figure*}
\includegraphics[width=160mm]{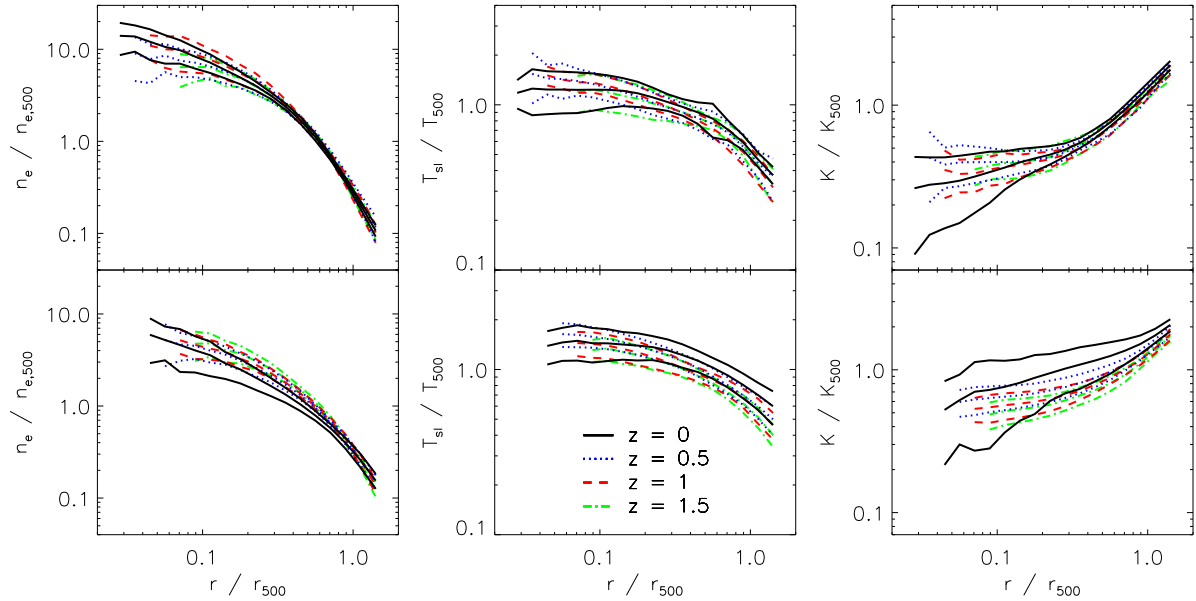}
\caption{Evolution of gas density, spectroscopic-like temperature and entropy profiles for clusters in the Millennium Gas FO run. The layout of the plots is identical to Figure \ref{fig:PCprofevln}.}
\label{fig:FOprofevln}
\end{figure*}

The evolution of the scaled gas density, spectroscopic-like temperature and entropy profiles of clusters in the FO run is illustrated in Figure \ref{fig:FOprofevln}. 

The main point to note from the scaled profiles of the $10$ hottest clusters is that they do not evolve with redshift. This means that the gas density, temperature and entropy scale in the same way as predicted by the self-similar model: $n_{\rm e}\propto E(z)^2$, $T_{\rm sl}\propto E(z)^{2/3}M^{2/3}$ and $K\propto M^{2/3}/E(z)^{2/3}$. Recall that the PC model predicts substantial evolution of scaled cluster profiles over the same redshift range, essentially because haloes are simply `recovering' from the extreme preheating at $z=4$. By contrast, energy feedback from galaxies is a continual process in the FO run, acting to reduce the gas density and increase the entropy in central regions. The fact that entropy is injected in such a way that the shape of the scaled cluster profiles does not change with redshift is presumably attributable to the self-regulatory nature of the feedback loop in the galaxy formation model underpinning our simulation. This behaviour is not peculiar to the most massive objects; we find that the scaled profiles of lower-mass systems evolve in a self-similar fashion too. We have also checked that cluster mergers have a negligible effect on the shape, dispersion and evolution of radial profiles by following the procedure outlined in the previous section.

Turning our attention to clusters with a temperature $3\ {\rm keV}\leq T_{\rm sl}\leq 4\ {\rm keV}$, we find that their scaled profiles do evolve, in the opposite sense to that predicted by the PC model. The explanation for this behaviour lies in the fact that we are considering a different set of clusters at each redshift. Consider the $z=1.5$ progenitors of clusters with a temperature $3\ {\rm keV}\leq T_{\rm sl}\leq 4\ {\rm keV}$ at $z=0$. These objects are less massive than clusters with a temperature in the same range at $z=1.5$. Accordingly, they will have been more affected by the non-gravitational heating from AGN, so the gas density will be lower and the entropy higher in central regions. Since the scaled profiles of individual clusters do not evolve in the FO run, these differences are preserved until $z=0$, so we see apparent signs of evolution when comparing the profiles of clusters with $3\ {\rm keV}\leq T_{\rm sl}\leq 4\ {\rm keV}$ at $z=0$ with clusters of the same temperature at higher redshift.

\subsection{Evolution of X-ray scaling relations from $z=1.5$ to $z=0$}
\label{sec:scalerelevln}

We have seen that feedback from galaxy formation leads to dramatically different evolution of cluster profiles than high-redshift preheating. It follows that we should see differences in the evolution of X-ray scaling laws too. We now examine whether the evolution of the \yxm, \tslm, \lxm\ and \lxtsl\ relations predicted by our FO simulation is compatible with observational data, and whether the data prefers this model over simple preheating. The datasets we use are the low-redshift REXCESS sample of PCA09, and the high-redshift sample of \citet[hereafter MJF08]{MJF08}. We choose to compare with the data of MJF08 for several reasons. First, this dataset is one of the largest high-redshift X-ray-selected cluster samples currently available, consisting of $115$ clusters observed with \emph{Chandra}. Second, it covers a broader redshift range ($0.1<z<1.3$) than any other existing large sample. Third, temperatures and bolometric luminosities were derived in the aperture $0<r\leq r_{500}$, using the same definition of $r_{500}$ as in this work. This agrees with the way in which these properties were calculated for our simulated clusters. Nevertheless, we must be careful not to over-interpret any comparison of our simulated data with observations since strong selection biases limit our ability to perform a statistically meaningful comparison. We discuss this further in section \ref{sec:obscomp} below.

To account for the evolution of X-ray observables, we define scaling relations by
\begin{equation}
\label{eq:genscalerelevln}
E(z)^nY=C(z)\left(\frac{X}{X_0}\right)^{\alpha},
\end{equation}
where all quantities are as in equation (\ref{eq:genscalerel}), except that the normalisation is now a function of redshift. For each relation, we fix the slope $\alpha$ to the $z=0$ value found previously (see Table \ref{tab:scalerelz0}), and compute the normalisation at each redshift by minimising $\chi^2$ in log space. The self-similar model predicts that the slope of each relation will be independent of redshift. In our simulations, we see small fluctuations in the slope with redshift, but there is no systematic variation, justifying our assumption of a fixed slope.

A power-law of the form
\begin{equation}
\label{eq:normevln}
C(z)=C_0(1+z)^{\beta},
\end{equation}
is then fit to the normalisation data to determine the parameters $C_0$ and $\beta$ (note that this may cause $C_0$ to change slightly from the $z=0$ value given in Table \ref{tab:scalerelz0}). Best-fitting parameters for each relation are listed in Table \ref{tab:normevln}. Since we have included the $E(z)^n$ factor in equation (\ref{eq:genscalerelevln}), then we would expect the slope $\beta$ to be zero if clusters do indeed evolve self-similarly. If $\beta<0$ or $\beta>0$, then we say there is negative or positive evolution, respectively. Note that some authors do not scale out the expected self-similar behaviour first, so their definition of negative/positive evolution has a different meaning to ours. This is one reason why care must be taken when comparing the results of different studies. 

\begin{table}
\caption{Best-fit parameters (with $1\sigma$ errors) for the evolution of the normalisation of the X-ray scaling relations predicted by each Millennium Gas simulation.}
\label{tab:normevln}
\begin{tabular}{@{}lcc}
\hline
Relation & $C_0$ & $\beta$ \\
\hline
GO simulation & & \\
\yxm & $4.222\pm 0.028$ & $-0.267\pm 0.011$ \\  
\tslm & $3.941\pm 0.031$ & $-0.335\pm 0.013$ \\ 
\lxm & $19.53\pm 0.65$ & $-0.243\pm 0.055$ \\ 
\lxtsl & $39.4\pm 1.4$ & $0.370\pm 0.058$ \\ 
\\
PC simulation & & \\
\yxm & $5.96\pm 0.23$ & $-0.330\pm 0.066$ \\  
\tslm & $6.317\pm 0.025$ & $0.0423\pm 0.0065$ \\ 
\lxm & $6.18\pm 0.42$ & $-0.90\pm 0.11$ \\ 
\lxtsl & $5.34\pm 0.52$ & $-1.77\pm 0.16$ \\
\\
FO simulation & & \\
\yxm & $5.683\pm 0.038$ & $0.054\pm 0.014$ \\  
\tslm & $6.180\pm 0.042$ & $-0.249\pm 0.014$ \\ 
\lxm & $6.334\pm 0.046$ & $0.748\pm 0.015$ \\ 
\lxtsl & $5.85\pm 0.14$ & $0.760\pm 0.050$ \\ 
\hline
\end{tabular}

\medskip
$C_0$ is the best-fitting normalisation, and $\beta$ is the best-fitting slope. $\beta$ characterises the evolution of the normalisation; see equations (\ref{eq:genscalerelevln}) and (\ref{eq:normevln}).
\end{table}

\subsubsection{The \yxm\ relation}

\begin{figure}
\includegraphics[width=85mm]{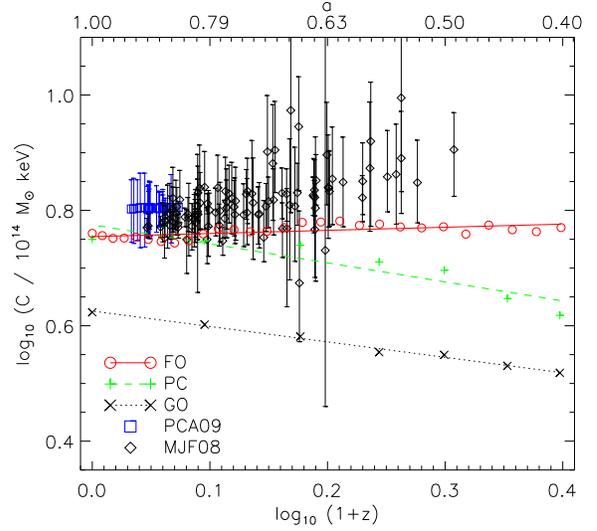}
\caption{Normalisation of the \yxm\ scaling relation as a function of redshift for each of the Millennium Gas simulations. Low-redshift observational data from REXCESS (PCA09) and the high-redshift data of MJF08 is shown for comparison. $1\sigma$ error bars are also plotted for the observational data.}
\label{fig:YM_evln}
\end{figure}

Figure \ref{fig:YM_evln} illustrates how the normalisation of the \yxm\ relation evolves in each of the Millennium Gas simulations. The observational data of PCA09 and MJF08 shown in the figure was plotted as follows. For each cluster in the two datasets, we computed $C(z)$ using equation (\ref{eq:genscalerelevln}), assuming $\alpha$ was fixed to the slope of the local \yxm\ relation of \citet{APP07}: $\alpha=1.825\pm 0.090$. Note that MJF08 adopt the same definition of \yx\ as used in both this work and REXCESS.

The normalisation of the \yxm\ relation evolves in a negative sense in the PC run, which can be understood as follows. Preheating drives significant amounts of gas beyond $r_{500}$ at high redshift, even in the most massive of haloes. Given some cluster at $z=0$, its gas mass at high redshift will thus be less in the PC run than expected from the pure gravitational heating scenario. On the other hand, Figure \ref{fig:PCprofevln} shows that the spectroscopic-like temperature increases with redshift relative to the self-similar prediction. Since \yx\ is the product of these two quantities, these opposing evolutionary trends will cancel each other out to some degree. However, the net effect is a more rapid decrease in \yx\ with redshift than anticipated from the self-similar model. Since the total mass of a preheated cluster will decrease with redshift at a rate akin to the self-similar prediction, this implies that any cluster on the PC \yxm\ relation at $z=0$ will follow a steeper trajectory towards the bottom-left of the \yxm\ plane than expected. The drop in normalisation with redshift will then be larger than in the self-similar model, so we see negative evolution.

Unlike preheating at high redshift, our model for feedback from galaxies produces a cluster population whose properties scale in a self-similar manner (see the top row of Figure \ref{fig:FOprofevln}). We would therefore expect to see no evolution of the normalisation of the FO \yxm\ relation, relative to the prediction of the self-similar model. From Figure \ref{fig:YM_evln} we see that this is almost the case; the value of $\beta$ obtained from fitting the normalisation data is close to zero. There is slight positive evolution which arises because the slope of the relation, $\alpha\approx 1.7$, is marginally steeper than the self-similar value. This effect is explained in full in Appendix \ref{app:A}. We note that the significant negative evolution seen in the GO model can be explained in the same way.

\subsubsection{The \tslm\ relation}

\begin{figure}
\includegraphics[width=85mm]{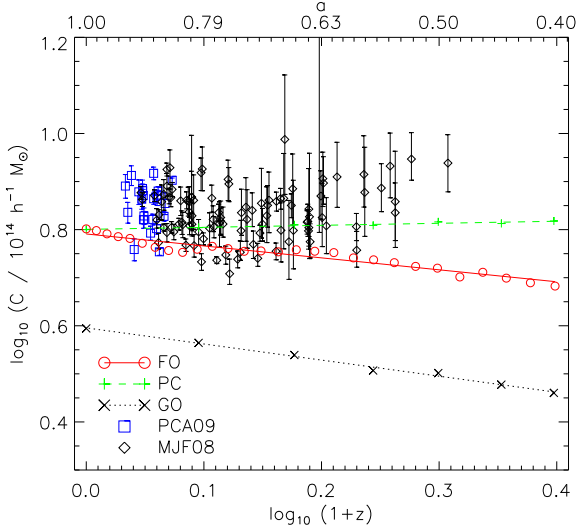}
\caption{Normalisation of the \tslm\ scaling relation as a function of redshift for each of the Millennium Gas simulations. We also display observational data from PCA09 and MJF08.}
\label{fig:TM_evln}
\end{figure}

The normalisation of the \tslm\ relations obtained from the GO, PC and FO runs is shown as a function of redshift in Figure \ref{fig:TM_evln}. Also shown is the observational data of PCA09 and MJF08, where we have assumed a fixed slope equal to that of the local REXCESS $T_{\rm spec}$-$M$ relation: $\alpha=0.633\pm 0.032$ (G.~W.~Pratt, priv. comm.).

The PC model predicts slightly positive evolution of the normalisation. If we take a particular object that lies on the \tslm\ relation at $z=0$, then we know from Figure \ref{fig:PCprofevln} that its temperature will decrease less rapidly with redshift than in the gravitational heating scenario. However, its total mass will decrease at a similar rate with redshift as it is dominated by the dark matter. Therefore, the cluster will move towards the bottom-left of the \tslm\ plane, following a shallower trajectory than expected from self-similar scaling arguments. It follows that, at some $z>0$, the normalisation will be higher than expected, i.e. positive evolution. Although the degree of evolution is small, recall that the slope of the PC \tslm\ relation is substantially shallower than the self-similar value (see Table \ref{tab:scalerelz0}). In Appendix \ref{app:A} we show how this can induce negative evolution, implying that the positive evolution predicted by the PC model is actually stronger than it appears.

The normalisation of the \tslm\ relation obtained from the FO run evolves in the opposite sense, being significantly negative. \emph{A priori} we would expect to find $\beta\approx 0$ in this model since entropy is injected in such a way that the profiles of individual clusters scale self-similarly (Figure \ref{fig:FOprofevln}). The cause of the apparent evolution is that the slope is considerably shallower than the self-similar prediction. Given this difference in slope, the argument presented in Appendix \ref{app:A} implies that individual clusters must decrease in mass by about a factor of three between $z=0$ and $z=1$ to account for the $\sim 20\%$ drop in normalisation seen in Figure \ref{fig:TM_evln}. We have checked that this is indeed the case in the FO simulation.

\subsubsection{The \lxm\ relation}

\begin{figure}
\includegraphics[width=85mm]{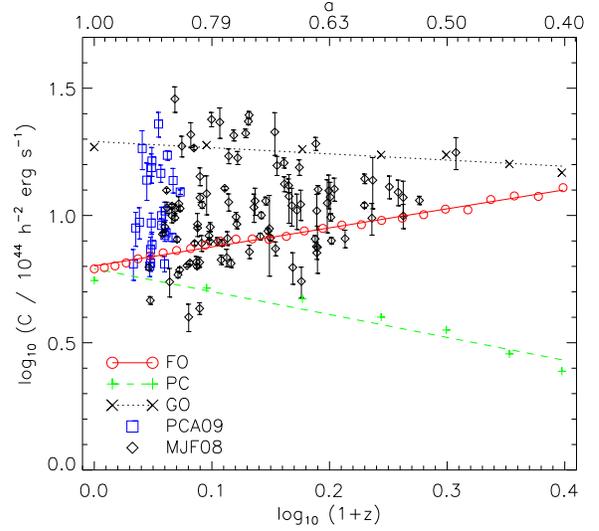}
\caption{Normalisation of the \lxm\ scaling relation as a function of redshift for each of the Millennium Gas simulations. For comparative purposes, we plot low and high-redshift observational data from PCA09 and MJF08, respectively. }
\label{fig:LM_evln}
\end{figure}

In Figure \ref{fig:LM_evln} we show how the normalisation of the \lxm\ relation evolves in each of our three simulations. The observational data of PCA09 and MJF08 is also plotted, assuming a fixed slope $\alpha=1.96\pm 0.11$, which is the slope of the low-redshift REXCESS \lxm\ relation.

The normalisation of the \lxm\ relation evolves in a negative manner in the PC run. To explain this, consider some preheated cluster that lies on the $z=0$ relation. Recall from Figure \ref{fig:PCprofevln} that the PC model predicts a drop in gas density, relative to the self-similar prediction, as redshift increases. Since the dominant contribution to the X-ray luminosity is the gas density in the Bremsstrahlung regime, then the luminosity at some higher redshift will be lower than predicted by the self-similar model. Given that the total mass of the cluster at this redshift will be close to the value expected from self-similar evolution, then we will see an apparent decrease in normalisation of the \lxm\ relation relative to the self-similar prediction. This corresponds to negative evolution. We would expect the evolution to appear stronger if the slope of the PC \lxm\ relation matched the self-similar value, rather than being considerably steeper.

The density and temperature profiles of clusters formed in the FO run evolve in a self-similar fashion, so the X-ray luminosity will also scale self-similarly. Since the growth rate of a cluster is governed primarily by the dark matter dynamics, we would thus expect the normalisation of the \lxm\ relation not to evolve once the predicted self-similar behaviour has been factored out. However, this will only be the case if the slope of the relation matches the self-similar value. In reality, feedback from SNe and AGN establishes a steeper slope, $\alpha\approx 1.8$. As discussed in Appendix \ref{app:A}, this departure from self-similarity leads to apparent positive evolution. From Figure \ref{fig:LM_evln}, we see that the normalisation of the FO \lxm\ relation does indeed evolve positively, with a $\sim 40\%$ increase in normalisation relative to the self-similar model between $z=0$ and $z=1$. This is consistent with the fact that the masses of clusters in the FO run decline by roughly a factor of three over this redshift range.

\subsubsection{The \lxtsl\ relation}

\begin{figure}
\includegraphics[width=85mm]{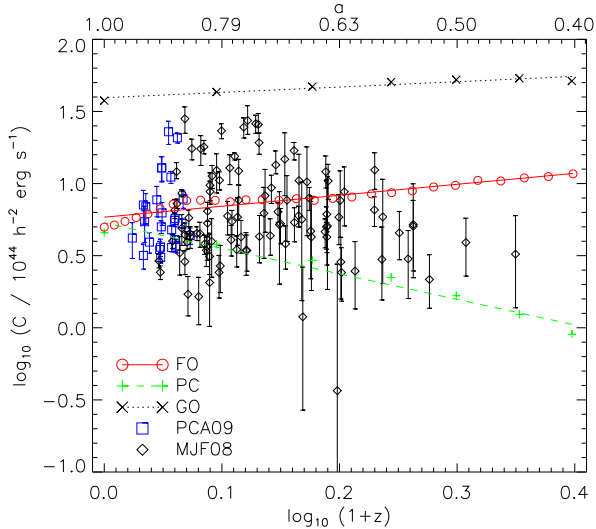}
\caption{Normalisation of the \lxtsl\ scaling relation as a function of redshift for each of the Millennium Gas simulations. Data from the observational studies of PCA09 and MJF08 is also shown.}
\label{fig:LT_evln}
\end{figure}

We show the normalisation of the three Millennium Gas \lxtsl\ relations as a function of redshift in Figure \ref{fig:LT_evln}. To plot the observational data of PCA09 and MJF08, we have fixed the slope to that of the local REXCESS relation: $\alpha=3.35\pm 0.32$.

The PC run predicts negative evolution of the \lxtsl\ relation as well, which is more pronounced than for the \lxm\ relation. This is because the temperature increases relative to the self-similar prediction with redshift (Figure \ref{fig:PCprofevln}), whereas the total mass decreases at a similar rate. Over a given redshift interval, the normalisation of the \lxtsl\ relation decreases more than that of the \lxm\ relation, so we see a larger drop in normalisation relative to the self-similar model, implying stronger negative evolution. As before, the evolution of the PC \lxtsl\ relation will have been tempered somewhat, because the slope is steeper than the self-similar value.

On the other hand, our model for feedback from galaxies leads to an \lxtsl\ relation with a positively-evolving normalisation. Given that the X-ray luminosity and spectroscopic-like temperature both scale self-similarly in this model, we should see no evolution relative to that expected from self-similar theory. However, it is evident from Figure \ref{fig:LT_evln} that, between $z=0$ and $z=1$, the normalisation of the \lxtsl\ relation increases by $\sim 50\%$ compared to the self-similar prediction. Again, this evolution arises because the slope of the \lxtsl\ relation obtained from the FO simulation, $\alpha\approx 3.3$, is much steeper than the self-similar prediction $\alpha=2$. With this difference in slope, the magnitude of the positive evolution can be readily explained by following the argument outlined in Appendix \ref{app:A}, using the fact that the temperatures of individual clusters in the FO run drop by about a factor of $2$ between $z=0$ and $z=1$.

We note that \citet{KDA07} find negative evolution of the \lxtsl\ relation using a fully self-consistent simulation with radiative cooling, star formation and supernova feedback. Their work is directly comparable to ours (same choice of overdensity, same temperature definition, no core excision, etc.), so this indicates that including AGN feedback changes the way in which the \lxtsl\ relation evolves, possibly because of the different redshift-dependence of the two feedback mechanisms. 

\subsubsection{Comparison with observations}
\label{sec:obscomp}

Energy feedback from galaxies leads to substantially different evolution of the \yxm, \tslm, \lxm\ and \lxtsl\ relations than uniform preheating. We now discuss which of the PC and FO models, if either, is preferred by the high-redshift data of MJF08. 

We begin by noting that, as in REXCESS, the masses of clusters in the sample of MJF08 were estimated from a \yxm\ relation. Since this relation was calibrated using clusters with hydrostatic mass estimates, the masses of their high-redshift clusters are also likely to be biased low by $\sim 10-20\%$. Therefore, the observational data points shown in Figures \ref{fig:TM_evln} and  \ref{fig:LM_evln} should all be shifted down by $\sim 5-10\%$ and $\sim 20-30\%$, respectively. 

Once we have applied this correction, we find that all four scaling relations obtained from the FO run evolve in a manner broadly consistent with the observational data at low to moderate redshifts, $z\lesssim 0.5$. In the case of the \lxm\ and \lxtsl\ relations, there are hints that the positive evolution predicted by our feedback model provides a better match to the data at these redshifts than the PC model, although the observed scatter is large. Both the PC and FO runs predict similar results for the \yxm\ and \tslm\ relations at $z\lesssim 0.5$, and it is not possible to distinguish between the two models with the observational data.

At higher redshift, $z\gtrsim 0.5$, the observational data for the \yxm\ and \lxm\ relations seems to follow an upward trend, consistent with the positive evolution expected from our feedback model. The data also suggests that the $T_{\rm spec}$-$M$ relation evolves in a positive sense at these redshifts, but in this case the PC model provides a better description of the observed evolution than the FO model. The PC model also predicts negative evolution of the \lxtsl\ relation for $z\gtrsim 0.5$, consistent with the observational data.

It is clear from this discussion that it is difficult to deduce whether the data of MJF08 favours our feedback model over simple preheating. For each relation, it seems as if the observed normalisation data cannot be well fit by a single power-law. For example, Figure \ref{fig:LT_evln} suggests that the evolution of the $L_{\rm X}$-$T_{\rm spec}$ relation is approximately self-similar (or possibly slightly positive) up until $z\sim 0.5$, then negative thereafter. This could be a signature of a change in the evolutionary behaviour of clusters that is not reproduced by any of our models or, more probably, it may simply be an artifact of selection effects instead. 

At low to moderate redshifts, most clusters in the heterogeneous sample of MJF08 come from samples based on the \emph{ROSAT All-Sky Survey} (RASS), which are wide and shallow. Their relatively high flux limit corresponds to an intermediate mass limit at low redshift, but a much higher mass limit at moderate redshift, thus falling on a steeper part of the mass function. Given the large scatter in the \lxm\ relation, this means that the number of objects scattered from the left to the right of the mass limit will grow relative to the number scattered in the other direction. This increasing bias towards luminous systems as we transition from low to moderate redshift may explain the `hump' in the observational data at $z\sim 0.3$ apparent in Figures \ref{fig:LM_evln} and \ref{fig:LT_evln}. Unfortunately, it is hard to quantify this effect since the sample of MJF08 is not cleanly selected, so the selection function is unknown.

At high redshift, we expect the data of MJF08 to be less affected by selection biases for two reasons. First, their high-redshift clusters come from narrow and deep samples, such as the \emph{Wide Angle ROSAT Pointed Survey} \citep{BVH07} and the \emph{$400$ Square Degree ROSAT PSPC Survey} \citep{HPE08}, whose lower flux limit corresponds to a high-redshift mass limit that falls on a flatter part of the mass function than the mass limit of RASS-based surveys at moderate redshift. This implies smaller bias given the same scatter in the \lxm\ relation as at lower redshift. Second, there should be less scatter about the mean \lxm\ relation at high redshift due to the absence of cool cores (e.g. \citealt{VBF07}), which would further reduce any bias. However, given the remaining uncertainties on the selection biases and the limited number of high-redshift clusters in the sample of MJF08, we cannot draw firm conclusions about the nature of the evolution at $z\gtrsim 0.5$.

To summarise, it is fair to say that the quality of current X-ray data is insufficient to place robust constraints on theoretical models. In addition to small numbers of high-redshift clusters and large measurement errors, existing heterogeneous samples are plagued by strong selection biases which can imitate genuine evolution. As demonstrated by \citet{PPA07}, correctly modelling the full source-selection process is crucial for measuring the evolution of scaling laws. In the near future, the XCS will provide a large sample of X-ray-selected clusters ($\sim 500$ objects) with $0<z\lesssim 1.5$ that have been analysed in a consistent manner across the full redshift baseline. The survey selection function will be well monitored, allowing selection effects to be properly included when analysing the evolution of the X-ray scaling relations. With such a dataset it will hopefully become possible to discriminate between theoretical models such as our PC and FO models, providing us with a valuable insight into cluster astrophysics.  

\section{Summary and conclusions}
\label{sec:conc}

In this paper, we set out to investigate the evolution of galaxy cluster X-ray scaling relations using numerical simulations. The evolution of scaling laws is crucial for constraining cosmological parameters with clusters surveys, and also offers a potentially powerful probe of the cooling and heating processes operating in clusters. Our main objective was to determine how including additional feedback from AGN in simulations affects the predicted evolution, and whether this is consistent with observations. Given that there is a substantial body of observational and theoretical evidence indicating that AGN are key in shaping the properties of galaxy clusters, it is clearly important to address this issue. However, all evolution studies to date have been based on simulations that only incorporate feedback from star formation.

The simulation we have used for our study -- the FO run -- is a new member of the Millennium Gas suite, presented for the first time here. The basic objective of the Millennium Gas Project is to add gas to the structures found in the original Millennium Simulation. The Millennium Gas simulations are ideal for studying the evolution of cluster properties, because their large volume ($500^3 h^{-3}\;{\rm Mpc}^3$) means that we can resolve statistically significant numbers of massive clusters at all relevant redshifts. Furthermore, we can follow the formation of the richest clusters, which are the objects actually observed at high redshift.

Feedback is implemented in our simulation using the hybrid scheme of \citet{SHT09}, where the energy input into the ICM by SNe and AGN is calculated from a SAM of galaxy formation. This guarantees that feedback originates from a realistic galaxy population, whereas fully self-consistent simulations often predict excessive star formation on cluster scales. 

Our main conclusions can be summarised as follows.

\begin{enumerate}
\item Non-gravitational heating from SNe and AGN in the FO run produces a $z=0$ cluster population whose radial temperature and entropy profiles broadly agree with those of NCC clusters in the REXCESS sample (PAP10). In particular, the temperature profiles are close to isothermal in the core, and the entropy profiles are significantly flatter in central regions than the theoretical $K\propto r^{1.1}$ scaling observed in cluster outskirts. However, it seems that the entropy of the gas has been raised too much in the core, compared to the observational data. None of our clusters exhibit a gentle drop in temperature at small cluster-centric radii or a steadily declining entropy profile, both of which are characteristic of CC systems. This is because gas cannot lose entropy via radiative cooling in our simulation. We note that fully self-consistent hydrodynamical simulations tend to suffer from the opposite problem, in the sense that radiative cooling leads to the over-production of cool cores. 

\item The \yxm, \tslm, \lxm\ and \lxtsl\ scaling relations obtained from the FO run at $z=0$ generally match the local REXCESS relations (PCA09), once we have accounted for the fact that the observed masses are likely to be biased low by $\sim 10-20\%$ due to the assumption of hydrostatic equilibrium. The exception is that we cannot explain the large scatter above the mean \lxm\ and $L_{\rm X}$-$T_{\rm spec}$ relations seen in the observational data. This is because the source of this scatter is highly X-ray luminous CC systems which are not formed in our simulation. 

\item A crude model of non-gravitational heating from astrophysical sources in which the ICM is preheated at $z=4$, rather than in response to galaxy formation, can produce a population of clusters whose $z=0$ properties closely resemble those of objects formed in the FO run. In fact, the two model predictions are so similar that they cannot be distinguished using high-quality local observations.

\item Density, temperature and entropy profiles of individual clusters in the FO run all evolve in a self-similar fashion from $z=1.5$ to $z=0$, although feedback from galaxies has modified their shape compared to that expected from pure gravitational heating. We suspect this is linked to the self-regulation of cooling and heating in the underlying model of galaxy formation.

\item The profiles of preheated clusters do not scale self-similarly. This is because the injection of entropy at high-redshift acts to remove gas from central cluster regions, lowering the gas density and increasing its temperature. Following preheating, the properties of the ICM can only be modified by gravitational processes, so the effect of the preheating is gradually erased and cluster profiles will eventually resemble those of clusters that have been subject to gravitational heating only. This `recovery' from preheating is what drives the apparent evolution of cluster profiles relative to the self-similar model. 

\item Feedback from galaxy formation in our FO model leads to positive evolution of the \yxm, \lxm\ and \lxtsl\ relations, and negative evolution of the \tslm\ relation. By contrast, preheating leads to scaling relations that evolve in the opposite sense. \citet{KDA07} also reported negative evolution of the \lxtsl\ relation using a simulation with a self-consistent stellar feedback scheme. This suggests that additional heating from AGN feedback changes the way in which scaling laws evolve, possibly because AGN heating is still important in cluster cores at low-redshift, long after the peak of star formation. We have investigated whether the evolution predicted by our feedback model is consistent with X-ray observations of high-redshift clusters. Unfortunately, the large samples of high-redshift clusters currently available are not cleanly selected, which is problematic since it may generate spurious evolution (e.g. \citealt{PPA07}). This is possibly why different observational studies give contradictory results. Consequently, we have not been able to decide whether our FO model provides a better description of reality than simple preheating. However, it is encouraging that the evolutionary behaviour predicted by the two models is distinct, particularly in the case of the \lxm\ and \lxtsl\ relations, so that we could potentially distinguish between them, and also other models (such as that of \citealt{KDA07}), when higher-quality data becomes available. As an example, the XCS will soon provide the largest ever sample of X-ray clusters selected with well-defined criteria, extending out to $z\approx 1.5$. Likewise, large high-redshift cluster samples are expected from SZ surveys currently underway. With such datasets, a rigorous comparison between theory and observation will become possible, so that we will be able to use the evolution of cluster scaling laws as an additional constraint on models of non-gravitational heating in clusters.
\end{enumerate}

The Millennium Gas FO simulation introduced here is the only existing simulation that is large enough to follow the evolution of significant numbers of massive clusters at reasonable resolution, while also attempting to include some of the main physical processes involved in cluster formation: star formation and feedback from both SNe and AGN. Although our feedback model can generally reproduce several key observational properties of clusters, at least for those without a cool core, it does have its limitations. In the future, we plan to enhance the hybrid model of \citet{SHT09} in two major ways.

First, we will adapt the model to follow the metal enrichment of intracluster gas by Type II and Type Ia SNe. This is important since radial abundance profiles derived from X-ray observations provide valuable constraints on the feedback mechanisms responsible for injecting metals into the diffuse phase (e.g. \citealt{DEM01,TKH04,VMM05}). \citet{COR06} have already used a similar approach to tackle this problem (see also \citealt{CTT08}). However, they neglected energy feedback from SNe and AGN in their hybrid model, which will have a significant impact on the way metals are distributed throughout the ICM.

Second, we aim to self-consistently incorporate radiative cooling into the model as well, rather than relying on the simple cooling recipes employed in SAMs. These recipes usually assume that haloes have a spherically symmetric isothermal gas distribution but, in general, neither of these assumptions will hold in hydrodynamical simulations. To circumvent this problem, we intend to fully couple SAMs to radiative simulations, so that the gas distribution in the simulation governs star formation, black hole growth and associated feedback in the SAM. This is a non-trivial task, requiring the simulation and SAM to be run simultaneously. With this modification we hope to be able to reproduce the roughly bimodal distribution of core entropies found in real clusters.  

\section*{Acknowledgements}

We are extremely grateful to B.~J.~Maughan and G.~W.~Pratt for supplying us with their observational data and offering helpful advice. We also thank V.~Springel for providing the merger tree software and G.~De Lucia for making the L-Galaxies semi-analytic model available to us. Simulations were performed using the HPC facility at Nottingham University and the Virgo Consortium Cosmology Machine at the Institute for Computational Cosmology, Durham. This work was supported by a Science and Technology Facilities Council rolling grant. OM is supported in part by the Thailand Research Fund and the Commission on Higher Education in Thailand (grant MRG5080314).

\bibliographystyle{mn2e} \bibliography{bibliography}

\appendix
\section{Mock evolution of X-ray scaling relations}
\label{app:A}

\begin{figure}
\includegraphics[width=85mm]{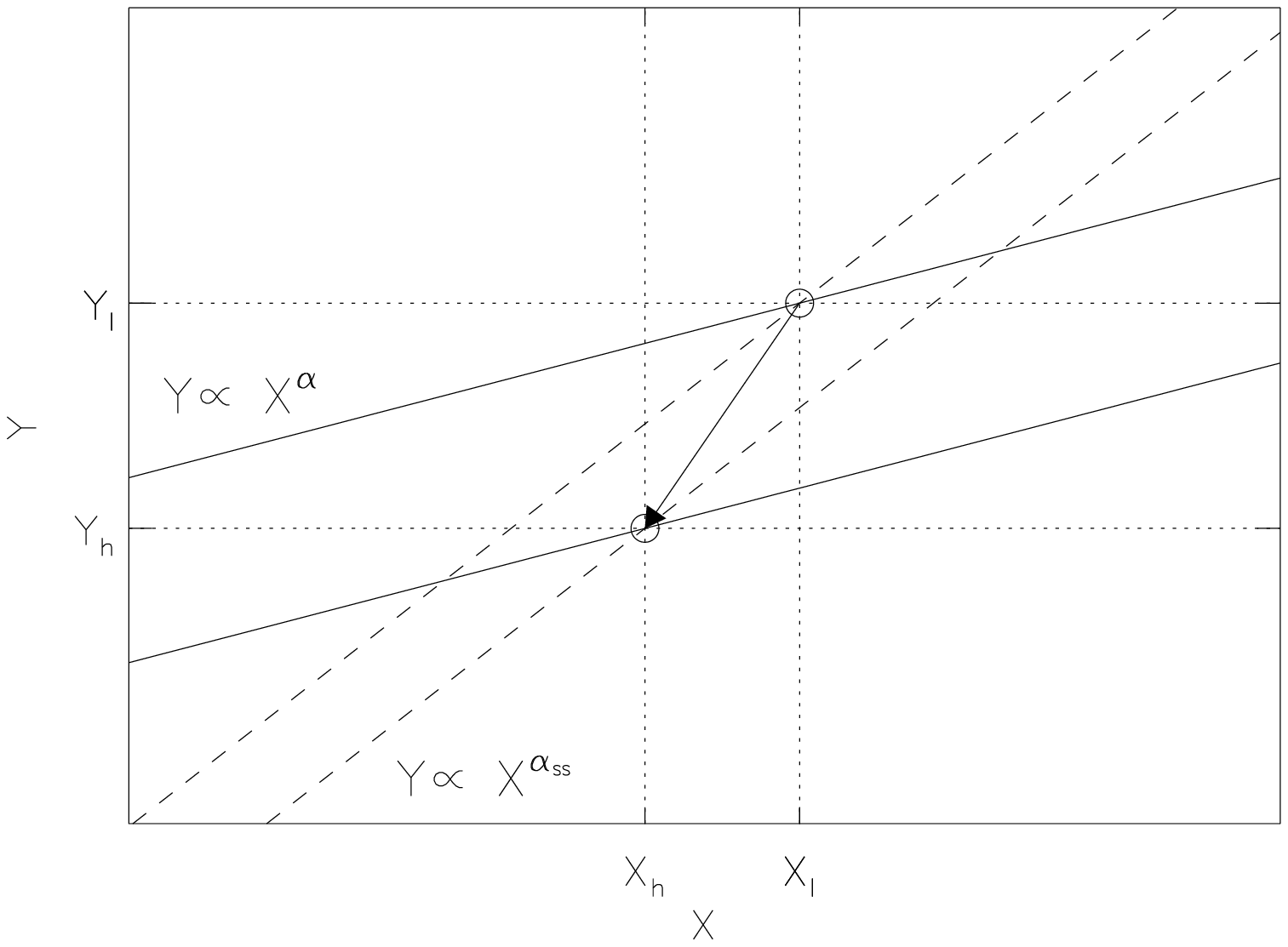}
\caption{Diagram to illustrate how evolution of a general X-ray scaling relation can be induced if the slope, $\alpha$, differs from the self-similar value $\alpha_{\rm SS}$. The arrow shows how the cluster  highlighted by an open circle evolves with increasing redshift. See text for mathematical details.}
\label{fig:slopeevln}
\end{figure}

In this Appendix we demonstrate how evolution of the scaling relations can arise if the slope is different to the self-similar prediction, even if cluster properties scale self-similarly themselves, as in our FO simulation. We present our argument in terms of the evolution of a general scaling relation of the form (\ref{eq:genscalerelevln}). A diagram of the situation under consideration is shown in Figure \ref{fig:slopeevln}.

Consider a cluster at $z_{\rm l}=0$ that is located at the point $(X_{\rm l},Y_{\rm l})$ on a $Y$-$X$ relation with a self-similar slope $\alpha_{\rm SS}$ and normalisation $C_0^{\rm SS}$, so
\begin{equation}
\label{eq:Yiss}
Y_{\rm l}=C_0^{\rm SS}\left(\frac{X_{\rm l}}{X_0}\right)^{\alpha_{\rm SS}}.
\end{equation}
Suppose the cluster evolves self-similarly until, at some redshift $z_{\rm h}>0$, it is located at the point $(X_{\rm h},Y_{\rm h})$, then
\begin{equation}
\label{eq:Yfss}
E(z_{\rm h})^n Y_{\rm h}=C_0^{\rm SS}\left(\frac{X_{\rm h}}{X_0}\right)^{\alpha_{\rm SS}}.
\end{equation}

In reality, we expect non-gravitational cooling and heating processes to alter the slope of the $Y$-$X$ relation from the self-similar prediction. Now suppose that our cluster was located at the same point, $(X_{\rm l},Y_{\rm l})$, at $z_{\rm l}$, but lay on a $Y$-$X$ relation with a slope $\alpha$ where, without loss of generality, we take $\alpha<\alpha_{\rm SS}$. If we assume self-similar scaling of cluster properties, as predicted by our feedback model, then the position of the cluster at $z_{\rm h}$ will again be $(X_{\rm h},Y_{\rm h})$. We can then write
\begin{equation}
\label{eq:Yi}
Y_{\rm l}=C_{\rm l}\left(\frac{X_{\rm l}}{X_0}\right)^{\alpha},
\end{equation}
and
\begin{equation}
\label{eq:Yf}
E(z_{\rm h})^n Y_{\rm h}=C_{\rm h}\left(\frac{X_{\rm h}}{X_0}\right)^{\alpha},
\end{equation}
where $C_{\rm l}$ and $C_{\rm h}$ are the normalisations of the $Y$-$X$ relation at $z_{\rm l}$ and $z_{\rm h}$, respectively. In the self-similar model we would have $C_{\rm h}/C_{\rm l}=1$. From equations (\ref{eq:Yi}) and (\ref{eq:Yf}) it follows that
\begin{equation}
\frac{C_{\rm h}}{C_{\rm l}}=E(z_{\rm h})^n\frac{Y_{\rm h}}{Y_{\rm l}}\left(\frac{X_{\rm h}}{X_{\rm l}}\right)^{-\alpha},
\end{equation}
but equations (\ref{eq:Yiss}) and (\ref{eq:Yfss}) imply
\begin{equation}
E(z_{\rm h})^n\frac{Y_{\rm h}}{Y_{\rm l}}=\left(\frac{X_{\rm h}}{X_{\rm l}}\right)^{\alpha_{\rm SS}},
\end{equation}
so
\begin{equation}
\frac{C_{\rm h}}{C_{\rm l}}=\left(\frac{X_{\rm h}}{X_{\rm l}}\right)^{\alpha_{\rm SS}-\alpha}.
\end{equation}
Since $X_{\rm h}<X_{\rm l}$ and $\alpha_{\rm SS}>\alpha$, then $C_{\rm h}/C_{\rm l}<1$, so we see a decrease in normalisation relative to the self-similar prediction with redshift, i.e. negative evolution. If the slope had been steeper than the self-similar value, $\alpha>\alpha_{\rm SS}$, then we would have seen apparent positive evolution.
\end{document}